\documentclass[11pt]{amsart}
\usepackage[margin=1in]{geometry}
\usepackage{lipsum}
\usepackage{amssymb, amsfonts, amsthm, mathrsfs, hyperref}
\usepackage{graphicx, xcolor}
\usepackage{dsfont}
\usepackage[all]{xy}
\usepackage{enumerate}
\usepackage[leqno]{amsmath}
\usepackage{comment}
\usepackage{dsfont}
\usepackage{subcaption} 
\usepackage{soul}

\theoremstyle{plain}
\newtheorem{theorem}{Theorem}%
\newtheorem{fact}[theorem]{Stylized Fact}
\newtheorem*{fact*}{Stylized Fact}

\newtheorem{proposition}[theorem]{Proposition}

\theoremstyle{remark}
\newtheorem{remark}[theorem]{Remark}

\numberwithin{equation}{section}

\allowdisplaybreaks[1]

\newcommand{\dd}{\mathrm{d}}

\DeclareMathOperator*{\sign}{sign}

\begin{document}

\title{Macroscopic properties of equity markets: stylized facts and portfolio performance}

\author{Steven Campbell}
\address{Columbia University}
\email{sc5314@columbia.edu}

\author{Qien Song}
\address{University of Toronto}
\email{johnsqe.song@alumni.utoronto.ca}

\author{Ting-Kam Leonard Wong}
\address{University of Toronto}
\email{tkl.wong@utoronto.ca}


\begin{abstract}
Macroscopic properties of equity markets affect the performance of active equity strategies but many are not adequately captured by conventional models of financial mathematics and econometrics. Using the CRSP Database of the US equity market, we study empirically several macroscopic properties defined in terms of market capitalizations and returns, and highlight a list of stylized facts and open questions motivated in part by stochastic portfolio theory. Additionally, we present a systematic backtest of the diversity-weighted portfolio under various configurations and study its performance in relation to macroscopic quantities. All of our results can be replicated using codes made available on our online repository. 

\end{abstract}

\maketitle{}

\section{Introduction} \label{sec:intro}
The development of quantitative finance and financial practice goes hand in hand with the analysis of financial data. Over the last few decades, academic researchers and practitioners have accumulated an enormous amount of empirical work on the temporal behaviours of individual and multiple asset prices at low and high frequencies, as well as cross-sectional properties in relation to macroeconomic, fundamental and statistical factors. Cont's 2001 paper \cite{C01} summarizes many by now classical {\it stylized facts} about individual assets. Also see \cite{BEM16, CLM97, F19, L19, T10} for textbook accounts of financial econometrics and empirical asset pricing theory. More recent developments include but are not limited to modelling of market microstructures and limit order books \cite{GPWMFH13}, rough volatility \cite{GJR18}, signature-based models \cite{CGS23}, econophysics \cite{chakraborti2011econophysics}, machine learning techniques in asset pricing \cite{GKX20}, energy markets \cite{CRS16, PT08} as well as cryptocurrencies and decentralized finance \cite{corbet2018exploring, ZWLS18}. These results not only improve our understanding of financial markets but also directly affect the design, implementation and risk management of trading strategies, especially for new markets and products. In each context, knowledge of stylized facts is essential since they highlight key properties that a realistic model should capture.  Despite the variety of markets, equity markets remain fundamental  because of their sheer sizes and primary functions of financing and trading of risks. 

In this paper we study empirically a collection of {\it macroscopic properties} of equity markets which are highly relevant in the management of large equity portfolios but are not adequately addressed by the aforementioned literature. In a nutshell, a macroscopic quantity or property of an equity market is one which depends on all or a majority of the stocks in the market. The most basic example is the total market capitalization, or overall performance, of the market which is well represented by a capitalization-weighted market index such as the S\&P 500. The index often serves as a benchmark for equity portfolios, both passive and active, and may be approximated by tradeable assets such as exchange-traded funds. Our terminology is motivated by statistical physics in which one considers macroscopic quantities such as volume, temperature and pressure.  To take our crude physical analogy a bit further, suppose we think of an equity market as a galaxy and each stock as a star. Then the market index is a proxy of the total mass, but there are other interesting macroscopic quantities, such as the shape, spiral and rotation of the galaxy, which affect and are affected by the stars, and are much less understood.\footnote{There are also other galaxies (markets) but they are beyond the scope of this paper.} From the viewpoint of, say, the manager of an equity portfolio which aims to outperform a capitalization-weighted benchmark, macroscopic properties are clearly relevant since they affect both the absolute performance of the strategy and the
relative performance with respect to the benchmark. For example, in \cite{AFG11, F02} it was shown empirically that changes in {\it market diversity}  (Section \ref{sec:diversity}), which measures the concentration of the market capitalization weights, correlate significantly with the relative return of actively managed large-cap portfolios. Macroscopic properties are also related to the overall stability of the market \cite{CHDPS07}. Viewing equity markets from the macroscopic perspective suggests many interesting questions which are not satisfactorily captured by conventional models of asset prices. The study of macroscopic features and their relationship with individual actions also arises in other areas in mathematical finance and applied probability, most notably in {\it mean-field games} \cite{carmona2018probabilistic,huang2007large,huang2006large,LL07}.

Our study is motivated by {\it stochastic portfolio theory} (SPT)  which was first pioneered by Fernholz \cite{F02, F20, FS82}. This mathematical theory provides a fresh perspective on some macroscopic properties of large equity markets, especially those defined in terms of market capitalizations and/or returns. In particular, market diversity and {\it intrinsic volatility} (relative volatility among the stocks, see Section \ref{sec:excess.growth.rate}), appear to be stable over long periods and, at least under suitable conditions, can be exploited by carefully chosen portfolios to outperform a capitalization-weighted benchmark. While the size factor has played an important role in empirical finance (see e.g. \cite{FF92}), SPT places great emphasis on how (relative) sizes simplify the modelling of the market, describe its stability, and allow us to analyze in depth the relative performance of a wide class of systematically rebalanced portfolios. Unfortunately, many results motivated by SPT remain largely inaccessible to non-experts due to their technical nature and possibly some restrictions imposed by the mathematical models (such as the interacting particle systems discussed in Section \ref{sec:capital.curve} and the focus on functionally generated portfolios). Also, the majority of papers in SPT focus on theoretical developments. Many important empirical questions, such as the joint modelling and prediction of diversity and volatility, and the calibration and diagnostics of high dimensional market models, are largely open.

Our paper, whose title is inspired by \cite{C01}, has three main objectives. First and foremost, we present a self-contained and unified empirical study of some macroscopic properties of the US equity market from 1962 to 2024, based on the CRSP US Stock Database, for a broad audience including financial mathematicians, econometricians, financial economists and portfolio managers. We substantiate, and sometimes modify, the ``folklore observations'' in SPT, thereby suggesting a collection of statistical and financial problems that we believe are of interest to the aforementioned communities. We summarize our main empirical findings which will be made more precise in the body of the paper:
\begin{itemize}
\item[{\bf 1.}] {\bf Capital distribution curve is stable.} The dependence of market capitalization weight on relative rank (by size) does not follow Zipf's law but has been stable over decades.
\item[{\bf 2.}] {\bf Market diversity is maintained through entrances and exits of stocks.} The capital distribution of a fixed collection of stocks tends to become more concentrated over time.
\item[{\bf 3.}] {\bf Relative (that is, cross-sectional) volatility clusters in time}. In particular, it exhibits positive serial correlation at multiple frequencies
\item[{\bf 4.}] {\bf Relative volatility tends to be large when market diversity is volatile and vice versa.}
\item[{\bf 5.}] {\bf Smaller stocks tend to be more volatile.}%
\item[{\bf 6.}] {\bf Smaller stocks have a higher tendency to be overtaken by smaller or new stocks.}
\item[{\bf 7.}] {\bf Rank switching occurs more intensely for small stocks.}
\end{itemize}
To facilitate replication of our empirical results and/or modification of the conventions and parameters adopted, we have made our codes publicly available.\footnote{See \url{https://github.com/stevenacampbell/Macroscopic-Properties-of-Equity-Markets}. The CRSP Data, which is licensed, must be obtained separately. We also describe how to download and clean the data.} Second, we perform a systematic empirical backtesting of various specifications of the {\it diversity-weighted portfolio} \cite{FGH98} which is a representative large cap, rule-based strategy of special interest in portfolio management in general, and in SPT in particular. We show how their performance, relative to a capitalization-weighted benchmark, depend on the (realized) macroscopic quantities, extending the empirical studies in \cite{RX20, TM21}. Our backtesting engine, which implements the algorithm in \cite{RX20} for long-only portfolios under proportional transaction costs, is also available on our repository and may be useful to other researchers and practitioners. Third, as a by-product of our exposition, we offer discussions on recent developments motivated by SPT, extending beyond those covered in the 2009 survey \cite{FK09}, again with the broader communities in mind.

The rest of the paper is organized as follows. We begin by describing, in Section \ref{sec:data}, the CRSP dataset of the US stock market and introduce some notations which will be used throughout the paper. Section \ref{sec:capdist.and.diversity} is concerned with the capital distribution curve and market diversity. In Section \ref{sec:excess.growth.rate} we study the market's excess growth rate which is a measure of intrinsic volatility and examine how it relates with market diversity. Section \ref{sec:rank.based.properties} considers how the behaviours of stocks depend systematically on their relative ranks. Our portfolio backtesting experiments are presented in Section \ref{sec:backtest}. Finally, in Section \ref{sec:conclusion} we summarize our findings and discuss several directions for future research. Although knowledge of SPT is not assumed for reading this paper, we include for completeness a brief but self-contained overview of its main ideas in Appendix \ref{sec:appendix.spt}. In Appendix \ref{sec:details} we recall the definition of semimartingale local time which motivates the quantities studied in Section \ref{sec:local.time.empirical}.

\section{Data} \label{sec:data}
\begin{figure}[t!]
\includegraphics[scale = 0.75]{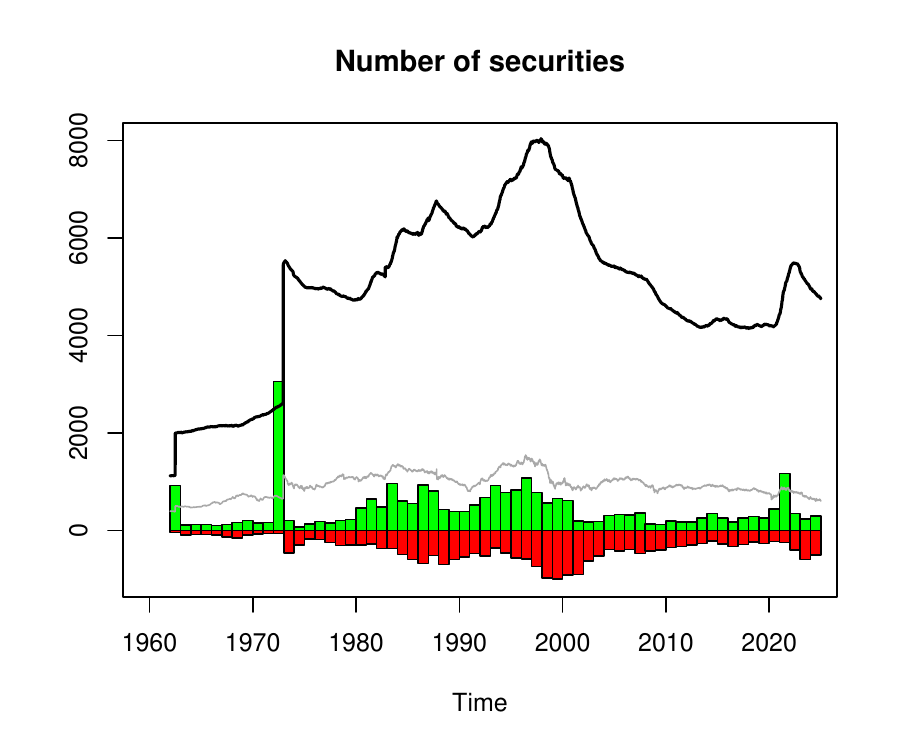}
\vspace{-0.2cm}
\caption{Overview of our CRSP data-set. Black (thick) series: Number of stocks in the full universe.  Grey (thin) series: Minimum number of stocks to cover at least $90\%$ of total market capitalization. Green (positive) bars: Number of new stocks. Red (negative) bars: Number of stocks delisted from the universe. The first two series are daily and the last two are yearly. The jumps in 1962 and 1972 are technical adjustments due to the inclusion of NYSE American and NASDAQ stocks.}
\label{fig:data}
\end{figure}

In this paper we focus on the US equity market for which detailed daily data is provided by the Center for Research in Securities Prices (CRSP).\footnote{The CRSP US Stock Databases can be accessed on the following website: \url{https://www.crsp.org/products/research-products/crsp-us-stock-databases}.} The CRSP database contains, among other attributes, the daily market capitalizations and returns of traded stocks listed on major US exchanges including NYSE, NYSE American, NASDAQ, NYSE Arca and BATS. It does not contain data for private companies that are not listed on stock exchanges. Dividends, corporate actions and delisting events are also included. Following \cite{RX20}, we implicitly treat the impact of corporate actions on portfolio performance through the CRSP fields \textsf{ret} and \textsf{dlret} corresponding to returns and delisting returns, respectively, that take into account these effects. We recommend Ruf's Python notebooks \cite{R24} for an accessible tutorial of the database. As in \cite{RX20}, we restrict to {\it common stocks} (more precisely, the securities with CRSP share codes (\texttt{shrcd}) 10, 11 and 12) and exclude other securities such as closed-end funds and REITs. While the data of NYSE stocks dates back to December 1925, stocks from NYSE American were included in CRSP starting July 1962 (and NASDAQ stocks in December 1972). NYSE Arca stocks were added in March 2006 but the effect was minor. For the purposes of this paper our data-set consists of the common stocks in the CRSP database from 1962-01-02 to 2024-12-31. Figure \ref{fig:data} provides an overview of our CRSP data-set. For each trading day $t$, let $\mathcal{A}_t$ be the set of all stocks which are traded on day $t$; we call it the {\it CRSP universe} on day $t$ and emphasize that it varies over time. We may think of each $i \in \mathcal{A}_t$ as the name or unique identifier of a listed company. As will be seen later, the inflow and outflow of stocks are highly relevant to the stylized features as well as the overall stability of the market. For $i \in \mathcal{A}_t$, let $X_i(t) > 0$ be the market capitalization of stock $i$ at the start of day $t$; it is, by definition, the product of the stock price and the number of outstanding shares. 

Despite the large number of stocks, the majority of market capital is concentrated in a relatively small number of the largest stocks. For example, throughout 2000--2024 the largest $1000$ stocks generally represent more than $90\%$ of the total market capitalization $\sum_{i \in \mathcal{A}_t} X_i(t)$. The concentration of capital among the stocks can be quantified using the concept of {\it market diversity} (Section \ref{sec:diversity}).

\section{Capital distribution curve and market diversity} \label{sec:capdist.and.diversity}

\begin{figure}[t!]
\includegraphics[scale = 0.6]{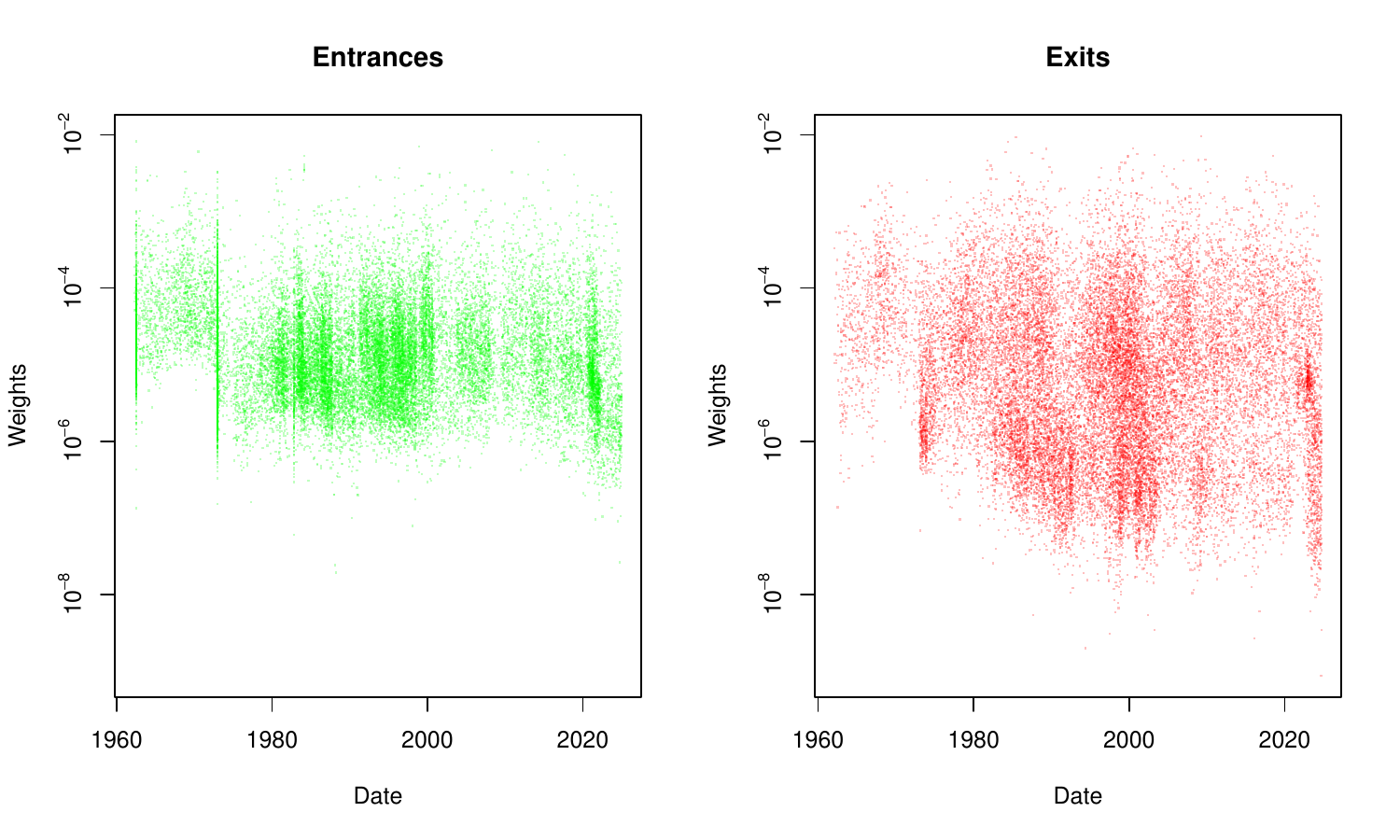}
\vspace{-0.55cm}
\caption{Inflow (left) and outflow (right) of the CRSP universe. Each green dot shows the time and capitalization weight (relative to the full CRSP universe) when a new stock enters the market. The red dots do the same for stocks which leave the market. The vertical green ``strips'' in 1962 and 1972 correspond to two artificial enlargements of the CRSP universe.}
\label{fig:Weights_Entrants_Exits_TS}
\end{figure}

\subsection{The capital distribution curve} \label{sec:capital.curve}
Let $\mathcal{I}_t \subset \mathcal{A}_t$ be a given collection of stocks on day $t$. For $i \in \mathcal{I}_t$, the {\it capitalization weight} of stock $i$ relative to $\mathcal{I}_t$ is defined by
\begin{equation} \label{eqn:market.weight}
\mu_i^{\mathcal{I}_t} = \mu_i^{\mathcal{I}_t} (t) := \frac{X_i(t)}{\sum_{j \in \mathcal{I}_t} X_j(t)}, \quad i \in \mathcal{I}_t.
\end{equation}
The capitalization weights define a probability vector with values in the (closed) {\it unit simplex}
\[
\mathcal{P}(\mathcal{I}_t) = \left\{ (\mu_i)_{i \in \mathcal{I}_t} : \mu_i \geq 0, \quad \sum_{i \in \mathcal{I}_t} \mu_i = 1  \right\},
\]
and can be regarded as the portfolio weights of a capitalization-weighted index. Nevertheless, naive implementation of this capitalization-weighted portfolio can be costly because of turnovers.\footnote{
 {\it Index tracking}, which is the main objective of passive portfolio management, is a topic of substantial practical importance, see e.g. \cite{canakgoz2009mixed, shu2020high} and the references therein.} Neglecting corporate events (such as dividends and delisting), a stock's capitalization weight increases (resp.~decreases) if it outperforms (resp.~underperforms) relative to the capitalization-weighted index.

Using the capitalization weights, we can visualize more accurately and vividly the evolution of the CRSP universe. In Figure \ref{fig:Weights_Entrants_Exits_TS}, we show with each green dot (left panel) the time and size of a stock when it is newly listed, and with each red dot (right panel) the same information for a stock which is delisted forever or for an extended period. We observe that both ``births'' (entrances) and ``deaths'' (exits) occur frequently\footnote{A number of market events can lead to entrances and exits of securities from the CRSP database. These include IPOs, bankruptcies, and mergers.} across a wide range of market capitalizations, yet their intensities appear to fluctuate over time. The weights at exit time are overall more spread out than the weights at entrance. An interesting statistical problem is to relate these fluctuations with the underlying economy such as business cycles.

The {\it capital distribution} relative to $\mathcal{I}_t$ is defined as the {\it ranked} probability vector 
\begin{equation} \label{eqn:capital.distribution}
\mu_{()}^{\mathcal{I}_t} = (\mu_{(1)}^{\mathcal{I}_t}, \mu_{(2)}^{\mathcal{I}_t}, \ldots, \mu_{(|\mathcal{I}_t|)}^{\mathcal{I}_t}),
\end{equation}
where $\mu_{(1)}^{\mathcal{I}_t} \geq \mu_{(2)}^{\mathcal{I}_t} \geq \cdots \geq \mu_{(|\mathcal{I}_t|)}^{\mathcal{I}_t}$ are the capitalization weights $(\mu_i^{\mathcal{I}_t})$ arranged in descending order. Since $\mu^{\mathcal{I}_t}_{(k)}$ decays rather quickly as $k$ increases, the capital distribution is usually visualized in log-log scale, i.e., $\log \mu_{(k)}^{\mathcal{I}_t}$ against $\log k$. We call the graph of $\log k \mapsto \log \mu_{(k)}^{\mathcal{I}_t}$ the {\it capital distribution curve} relative to the given universe (we use base $10$ for plotting purposes). When $\mathcal{I}_t = \mathcal{A}_t$ is the full universe, we call it the {\it full} capital distribution curve.

\begin{figure}[t!]
\includegraphics[scale = 0.7]{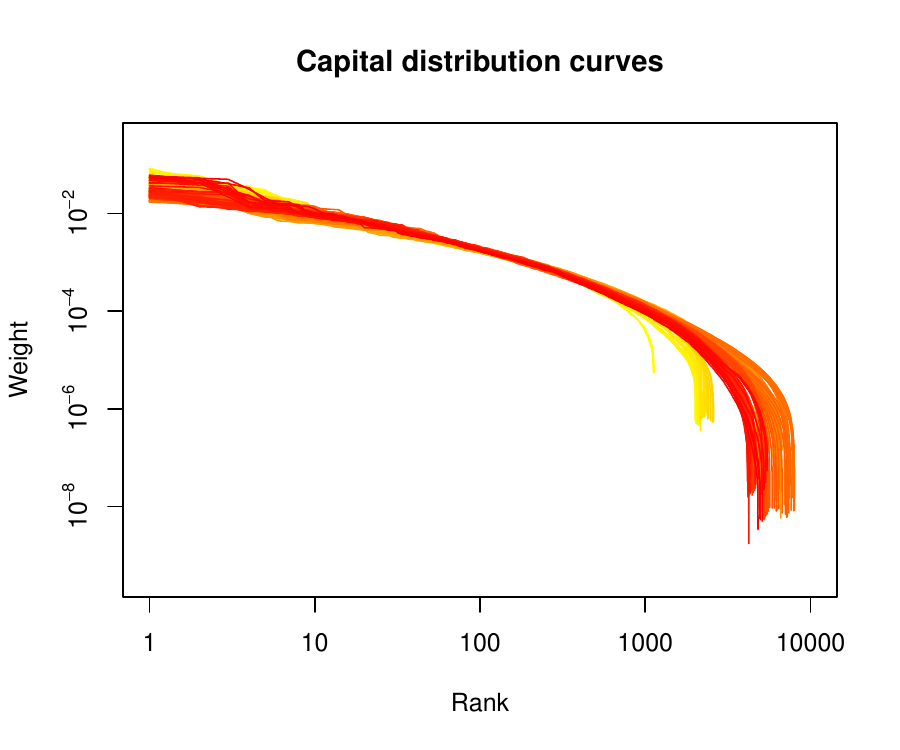}
\vspace{-0.2cm}
\caption{Capital distribution curves from 1962--2024 of the full CRSP universe. The curves are colored from yellow (more distant) to red (more recent).}
\label{fig:full_capdist}
\end{figure}

\begin{figure}[t!]
\includegraphics[scale = 0.25]{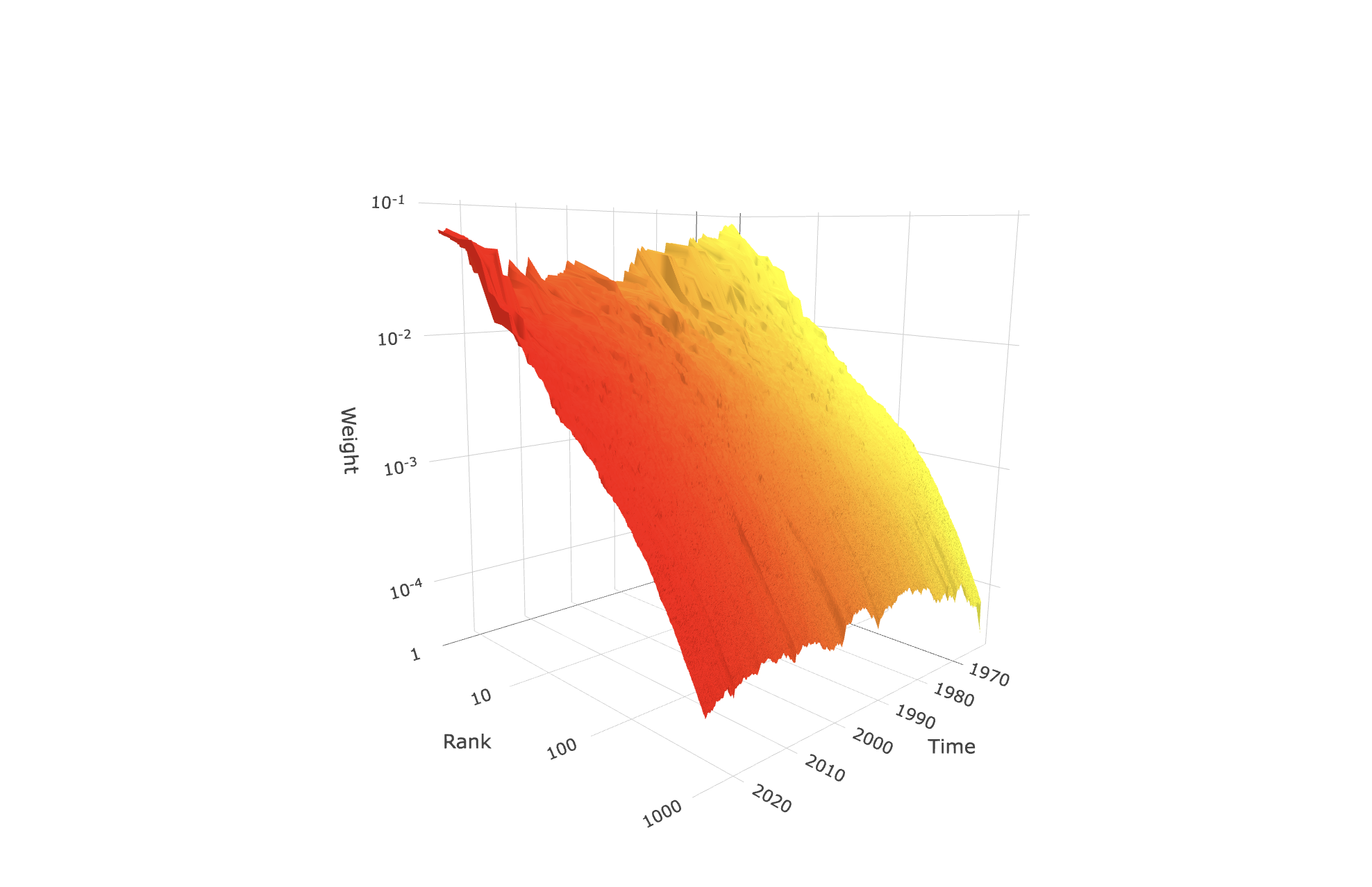}
\vspace{-0.1cm}
\caption{Capital distribution curves of the top $1000$ stocks from 1962--2024, visualized collectively as a surface. Colored from yellow (more distant) to red (more recent).}
\label{fig:top_capdist}
\end{figure}

In Figure \ref{fig:full_capdist} we plot the full capital distribution curve, which is one of the most iconic objects in SPT (see e.g.~\cite[Chapter 5]{F02}), from 1962 to 2024. The evolving dimension of $\mathcal{A}_t$ is inconvenient for conventional statistical methods. To avoid this issue, one may restrict to the subuniverse $\mathcal{A}_t^K$ consisting of the largest $K$ stocks of $\mathcal{A}_t$;\footnote{Exact ties occur extremely rarely, if at all, and can be resolved arbitrarily.} this amounts to renormalizing $(\mu^{\mathcal{A}_t}_{(1)}(t), \ldots, \mu^{\mathcal{A}_t}_{(K)}(t))$ to be a probability vector. The capital distribution curves relative to $\mathcal{A}^K_t$ with $K = 1000$ are visualized collectively in Figure \ref{fig:top_capdist} as a {\it surface} over the (log rank)-date plane. A glance at  Figures \ref{fig:full_capdist} and \ref{fig:top_capdist}  should convince the reader that the capital distribution curve is a rather delicate object to model faithfully - either directly or through a {\it market model} which specifies the joint dynamics of all stocks. Here, we observe that the left end of the curve is more volatile (in log-scale) than the right end, even though smaller stocks are generally more volatile in absolute terms (see \cite[Section 5.3]{CW22b} and Section \ref{sec:rank.based.properties}). Nevertheless, the overall shapes of the capital distribution curves appear to be quite stable. Thus we state our first stylized fact, somewhat loosely, as follows:

\begin{fact} \label{fact:1}
The overall shape of the capital distribution curve is stable. In particular, the full capital distribution curve is generally concave and decays rapidly on the right end.
\end{fact}

The apparent long-term stability, which we distinguish from the concept of (strict or second order) {\it stationarity} in time series analysis, of the shape of the capital distribution curve is striking and calls for explanation. As noted in \cite[pages 101--103]{D19}, this stability is neglected by standard multivariate financial models. To take a simple example, suppose stocks followed a multivariate Black-Scholes model as in the classical Merton problem \cite{M69}. Then the capital distribution would eventually degenerate into a point mass by the law of large numbers. If the capital distribution followed approximately a {\it power law} (or {\it Pareto distribution}) with exponent $\alpha > 0$, i.e., $\mu^{\mathcal{A}_t}_{(k)} \approx \frac{1}{Z_t} k^{-\alpha}$ where $Z_t$ is a normalization constant, the capital distribution curve would look approximately like a straight line with slope $-\alpha$. The special case $\alpha = 1$ is frequently called {\it Zipf's law}. While power laws are ubiquitous in many rank–size distributions in economics (see e.g.~\cite{FF20, G99, N05, SMS09} and the references therein) and {\it complex systems} \cite{thurner2018introduction}, it does not hold for the full capital distribution curve but may be a reasonable first approximation. We also remark that because the capital distribution curve is plotted in log-log scale, a visually small change of the curve, especially on its right end, corresponds to large movements in the capitalization weights.      

A possible interpretation of the stability of shape is that the capital distribution curve, which may be regarded as a high dimensional time series, can be well approximated by a low dimensional object which is stable in some sense. This suggests the use of dimension reduction. Using {\it convex principal component analysis \cite{BGKL17}} which can incorporate the monotone constraint of the capital distribution \eqref{eqn:capital.distribution}, the first and the last author showed in \cite[Section 6.3]{CW22b} that by restricting to, say, the largest $1000$ stocks (which differ from day to day) and using the {\it Aitchison geometry} \cite{EPMB03} on the (ranked) unit simplex, about $80\%$ of the variation of the curve can be captured by the first two convex principal components. Moreover, movements along the first convex principal component correlate strongly with {\it market diversity} whose behaviours are examined in Section \ref{sec:diversity}. Nevertheless, dimension reduction methods generally cannot capture idiosyncratic behaviours of the stocks, especially the largest ones which have significant influence on the market. 

Consider the problem of constructing stochastic models of market capitalizations which are capable of reproducing some features of the observed capital distribution curve (and ideally other behaviours of the market). A reasonably realistic model not only improves our understanding of the market but also serves as a {\it scenario generator} for optimizing and backtesting trading and risk management strategies. To enforce stability, a far-reaching idea, first proposed by Fernholz in \cite[Section 5.5]{F02} based on empirical observations, is to let the dynamics of a stock depend on its current rank. This modelling assumption motivated novel {\it rank-based diffusions}, and related processes such as {\it volatility-stabilized processes} and {\it polynomial processes}, for which a substantial probabilistic literature has accumulated in the last two decades (see e.g.~\cite{BFK05, CP10, cuchiero2019polynomial, cuchiero2018polynomial, FK05, FF20, IKS13,  IPBIF11, IL21, KS19, P11, ST17, SY24} and their references). Rank-based models have also been applied in economics to model the distribution of wealth and wealth mobility, see e.g.~\cite{BBF22, F16}. While it is beyond the scope of this paper to describe the variety of the mathematical results obtained, we will explain the main ideas. 

For theoretical tractability, one typically considers an equity market consisting of a fixed (finite or countably infinite) collection of stocks; in particular, IPOs and delisting events are neglected even though they are prominent as seen in Figure \ref{fig:data}.\footnote{A notable exception is \cite{karatzas2016diverse} which studied models with splits and mergers.} Nevertheless, by considering the largest $K \leq n$ stocks of the system, one can obtain a simulated universe which evolves over time via rank switchings.\footnote{The subsystem consisting of the largest $K$ stock is called an {\it open market} in \cite{KK21}.} As a basic example of a rank-based diffusion model, we mention the {\it generalized Atlas model}, introduced in \cite{BFK05}, under which the market capitalizations $X_1(t), \ldots, X_n(t)$ of $n$ stocks are modelled in continuous time as a Markov diffusion process satisfying the following system of stochastic differential equations:
\begin{equation} \label{eqn:rank.based.diffusion}
\dd \log X_i(t) = \sum_{k = 1}^n \gamma_k \mathds{1}_{\{\mathsf{rank}_i(t) = k\}} \dd t + \sum_{k = 1}^n \sigma_k \mathds{1}_{\{\mathsf{rank}_i(t) = k\}} \dd W_i(t), \quad i = 1, \ldots, n,
\end{equation}
where $\mathsf{rank}_i(t)$ is the rank of stock $i$ at time $t$, $\gamma_k$ and $\sigma_k$ are suitable {\it rank-based} constants and $W_1, \ldots, W_n$ are independent Brownian motions. In words, \eqref{eqn:rank.based.diffusion} says that the log market capitalization, $\log X_i$, of stock $i$ has drift $\gamma_k$ and volatility $\sigma_k$ when it has rank $k$ at time $t$. Focusing on rank-based behaviours limits the number of parameters in a parsimonious fashion as these features are typically more stable and eliminate the need for tracking the dependencies in a changing collection of securities. For example, the system \eqref{eqn:rank.based.diffusion} is specified by $2n$ parameters while an unrestricted covariance matrix based on the security names requires $\frac{1}{2}n(n+1)$ with $n$ increasing as new securities are introduced. In Section \ref{sec:rank.based.properties} we report some strong empirical evidence of rank dependence. {\it Interacting particle systems} such as \eqref{eqn:rank.based.diffusion} are not straightforward to analyze since the coefficients are discontinuous in the state variable.\footnote{In \eqref{eqn:rank.based.diffusion} the coefficients are piecewise constant. Existence of a weak solution follows from the theory in \cite{BP87}.} Also, the dynamics of the {\it ranked} market weights $\mu_{(k)}(t)$ and the {\it gaps} $\log \mu_{(k)}(t) - \log \mu_{(k+1)}(t)$ involve delicate concepts such as local times, collisions, as well as, reflections over polyhedral boundaries. Under suitable structural conditions on the parameters, it can be shown that a process such as \eqref{eqn:rank.based.diffusion} is ergodic and has a unique stationary distribution, under which the (expectation of the) capital-distribution curve is qualitatively similar to the ones shown in Figure \ref{fig:full_capdist}. They also capture some other rank-based properties such as the {\it higher volatility of smaller stocks} (Section \ref{sec:rank.volatility}). 
Nevertheless, we observe that most, if not all, results in this literature rely on ergodicity in the limit $t \rightarrow \infty$  and the stationary distribution. Consequently, these models need not, and usually do not, capture (qualitatively or quantitatively) short and medium term behaviours of the observed equity market. %

The model in \eqref{eqn:rank.based.diffusion} is called a {\it first order} model since the coefficients depend only on the ranks. A {\it second order} or {\it hybrid} model, as in \cite{FIK13, IPBIF11, IL21} involves also {\it name-based} coefficients and can incorporate idiosyncratic features. %
Calibration of ranked-based and volatility-stabilized market models to market data, as done in \cite{BBF22, F02, FK05, IKS13, IPBIF11, IL23}, is typically based on ad hoc methods analogous to moment matching and elementary estimates of growth rates and volatility coefficients (with smoothing); accuracy of the estimates and diagnostic checks are not addressed. A natural problem is to improve these models, their calibration and predictive power, possibly using tools in machine learning such as {\it neural stochastic differential equations} (see e.g.~\cite{LWCD20}). Moreover, market models such as \eqref{eqn:rank.based.diffusion} are {\it exogenous}; they do not explain why and how the stated dynamics emerge from interactions of the investors and the underlying economy. An ambitious goal is to come up with economically sound models of the financial market under which the observed macroscopic features emerge. Ideas from {\it evolutionary finance} \cite{BE92, EHS09}, {\it mean-field games} \cite{CD18, ichiba2020relative, kakeu2022size, LL07} and {\it complex systems} \cite{thurner2018introduction} may be relevant in modelling the dynamics of firms and interactions among investors and/or their strategies.

\subsection{Market diversity} \label{sec:diversity}
{\it Diversity indexes} are used in various fields to quantify the complexity or concentration of a system or population consisting of multiple ``species''. Examples include the Gini index for wealth inequality in economics, Hill numbers for biodiversity in ecology, and entropy as a measure of randomness in information theory and statistical physics. A comprehensive mathematical treatment can be found in \cite{L21}.\footnote{We thank Martin Larsson for pointing us to this interesting reference.} For an equity market, a natural idea is to quantify the diversity of the capitalization weights $ \mu^{\mathcal{I}_t}(t) =  \big(\mu_i^{\mathcal{I}_t} (t), i \in \mathcal{I}_t \big)$ as a probability vector indexed by the names, or the capital distribution $\mu_{()}^{\mathcal{I}_t}(t)$ as a probability vector indexed by the ranks,  by an entropy-like measure. In this context, the first paper we are aware of is \cite{F99}, in which the Shannon entropy was used. Unless otherwise stated, in this paper we measure market diversity of a collection $\mathcal{I}_t$ of stocks by the {\it Shannon entropy} of $\mu^{\mathcal{I}_t}(t)$ defined by
\begin{equation} \label{eqn:Shannon.entropy}
{\bf H}( \mu^{\mathcal{I}_t}(t) ) := -\sum_{i \in \mathcal{I}_t} \mu_i^{\mathcal{I}_t}(t) \log \mu_i^{\mathcal{I}_t}(t).
\end{equation}
By symmetry of the entropy we have ${\bf H}(\mu^{\mathcal{I}}(t)) = {\bf H}(\mu^{\mathcal{I}_t}_{()}(t)$). For a universe with a {\it fixed size} $|\mathcal{I}_t|$, ${\bf H}( \mu(t))$ is maximized (with value $\log |\mathcal{I}_t|$) when $\mu(t)$ is the uniform distribution, i.e., $\mu_i^{\mathcal{I}_t}(t) = \frac{1}{|\mathcal{I}_t|}$, and is minimized (with value $0$) when some stock completely dominates the universe, i.e., $\mu_i^{\mathcal{I}_t}(t) = 1$ for some $i$ and $\mu_j^{\mathcal{I}_t}(t) = 0$ for $j \neq i$. Apart from the Shannon entropy, other diversity measures (see \cite[Section 3.4]{F02}) may be used and produce qualitatively similar results; a specific example is the parameterized diversity measure ${\bf D}_{p}$ defined by
\begin{equation} \label{eqn:Fernholz.diversity.def}
{\bf D}_{p}( \mu^{\mathcal{I}_t}(t) ) := \Big( \sum_{i \in \mathcal{I}_t} \big(\mu_i^{\mathcal{I}_t}(t)\big)^{p} \Big)^{\frac{1}{p}},
\end{equation}
where $p \in (0, 1)$ is a tuning parameter. It is closely related to the {\it R\'{e}nyi entropy} (see \cite[Proposition 2]{W19}) and the {\it diversity-weighted portfolio} \eqref{eqn:diversity.weighted.portfolio} whose performance will be backtested, under various specifications, in Section \ref{sec:backtest}. 

\begin{remark}\label{rmk:div.and.perf}
Under the classical setting in SPT, the change in $\log {\bf D}_{p}$ is a major component of the relative performance of the diversity-weighted portfolio with respect to the market portfolio, in the short to medium term. See the ``master formula" \eqref{eqn:diweighted.portfolio} which holds under suitable modeling assumptions. More generally, in \cite{AFG11} (also see \cite{FG99, FGH98}) it was shown that the diversity of the S\&P 500  -- which corresponds roughly to $\mathcal{A}_t^{500}$ -- correlates significantly with the average relative performance of large-cap managers. Additional empirical results will be reported in Section \ref{sec:backtest}.
\end{remark} 

Being a summary statistic of the capital distribution, market diversity is related to the shape of the capital distribution curve. Roughly, diversity is higher when the curve is ``flatter'' and is lower when the curve is ``steeper''. In \cite{CW22b} we used convex PCA to make this idea precise: market diversity correlates strongly with the projection of the capital distribution curve onto the first convex principal component (analogous to how the the first and second principal components of yield curves can usually be interpreted in terms of the ``level'' and the ``slope'' of the curve). Since the capital distribution curve is ``stable'' by Stylized Fact \ref{fact:1}, it is natural to expect that market diversity has a tendency to decrease when it is ``too high'' and a tendency to increase when it is ``too low''. This idea is often expressed in the following way (cf.~the industry whitepaper \cite{F05}, the discussion in \cite[Section 7]{FK09} and the model choices in \cite{AFF07}):

\begin{fact*}[Folklore]
Market diversity (quantified by, say, the Shannon entropy) is mean-reverting.
\end{fact*}

\noindent
In the following we show that the behaviours of market diversity depend crucially on the {\it choice of the universe} $\mathcal{I}_t$. On the other hand, that diversity is ``mean reverting'' is subject to doubt. In the strictest sense, mean reversion means diversity is well approximated by an autoregressive or Ornstein-Uhlenbeck process about a fixed (or possibly varying) equilibrium level. While our empirical results do not provide conclusive evidence of this claim, long-term probabilistic behaviours of diversity and other observables can be analyzed mathematically for rank-based systems which generalize \eqref{eqn:rank.based.diffusion}; see \cite{MSZ19}.

\begin{figure}[t!]
\includegraphics[scale = 0.5]{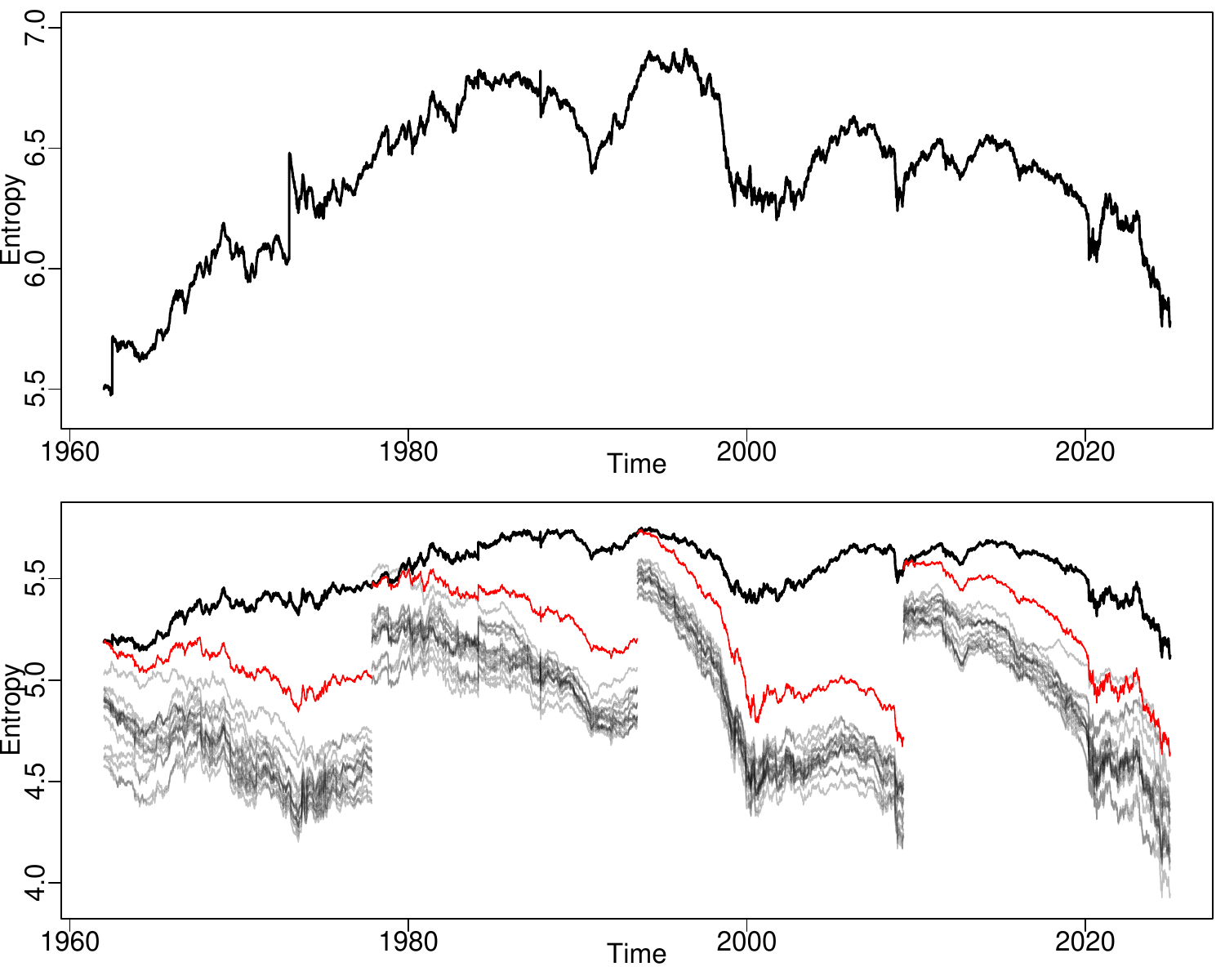}
\caption{Diversities (entropies) of several universes. Top: Full CRSP universe. Bottom: (i) (black) largest $500$ stocks, renewed daily; (ii) (red) largest $500$ stocks, renewed at the start of each subinterval; (iii) (grey) 25 randomly chosen collections, each containing $500$ stocks, renewed for each subinterval.}  
\label{fig:diversity}
\end{figure}

In the top panel of Figure \ref{fig:diversity} we show the time series of the diversity  ${\bf H}(\mu^{\mathcal{A}_t}(t))$ of the {\it full} CRSP universe. Observe that ${\bf H}(\mu^{\mathcal{A}_t}(t))$ correlates to some extent with $|\mathcal{A}_t|$, the total number of stocks (see Figure \ref{fig:data}). This is due to a {\it size effect} in the entropy (and other reasonable measures of diversity) \cite{L21}. For example, for $K \geq 1$ and $\alpha > 0$, consider the ranked probability vector $p = (p_k)_{k = 1}^K$ where $p_k = \frac{1}{Z_{K, \alpha}}k^{-\alpha}$ follows a power law with exponent $\alpha$ (here $Z_{K, \alpha} = \sum_{k = 1}^K k^{-\alpha}$ is the normalizing constant). Then one can verify that for $\alpha > 0$ fixed, ${\bf H}(p)$ is monotonically increasing in $K$. On the other hand, for $\alpha > 0$, the {\it normalized entropy} ${\bf H}(p) / \log K$ (which takes values in $[0, 1]$) is decreasing in $K$.\footnote{Also, the size effect may outweigh an increase in overall concentration. For example, if $p_k = {Z_{K, \alpha}}k^{-\alpha}$, $1 \leq k \leq K$ and $p_k' = {Z_{K', \alpha'}}k^{-\alpha'}$, $1 \leq k \leq K'$, are two Pareto distributions where $K' > K$ and $\alpha' > \alpha$, it is possible to have ${\bf H}(p') > {\bf H}(p)$. So comparing the diversities of two markets with different sizes is not an immediate task.} Thus it is difficult to argue that ${\bf H}(\mu^\mathcal{A}_t(t))$ has a fixed long term equilibrium value to which diversity is reverting. Intuitively, we might expect that a diverse market is more resilient to sudden shocks and catastrophic events, but the relation between diversity and systemic risk is, to the best of our knowledge, largely open. Also see \cite{JL21} for a comparative study of the systemic risks of international equity markets using a skewness-based measure. 

Observe that in 2023 and 2024, even after the COVID-19 recession, the diversity of $\mathcal{A}_t$ dropped to the lowest level since the 1970s. The  recent decrease in market diversity, largely attributable to seven mega-capitalization stocks, dubbed the ``Magnificent Seven'', is affecting active managers' stock selection behaviour. The {\it active-share ratio}, which measures how active portfolios deviate from the benchmark S\&P 500, is at its lowest level since 2013 \cite{fundprosactive}, indicating that active portfolio managers are choosing to mimic the benchmark more because missing out on any of the top-weighted stocks can lead to underperformance. If this trend continues, the (left end of the) capital distribution curve will look slightly steeper than before.

Next consider the behaviour of ${\bf H}(\mu^{\mathcal{I}_t}(t))$ for suitably chosen subuniverses $\mathcal{I}_t$ with $|\mathcal{I}_t| \leq K$; this makes it more straightforward to interpret the change in diversity over time. In the bottom panel of Figure \ref{fig:diversity} we consider three cases where $K = 500$:
{\renewcommand{\theenumi}{(\roman{enumi})}
\begin{enumerate}
\item[\refstepcounter{enumi}\theenumi\label{blk.srs}] (Black series) $\mathcal{I}_t = \mathcal{A}^K_t$ is the largest $K$ stocks on day $t$ renewed every day. Now diversity fluctuates in a much tighter interval compared to the case of full universe, but even with this restriction it is difficult to argue the existence of a fixed equilibrium value. Note that during the the period 1962--1972 the CRSP universe was artificially enlarged twice due to the addition of NYSE American and NASDAQ stocks. The diversity of $\mathcal{A}^K_t$ increased rather steadily in this period. We also observe that diversity appears to exhibit short to medium term trends or momentum. Also, several sudden drops in market diversity are associated with market crashes such as the financial crisis in 2008 and COVID-19 in 2020. 
\item[\refstepcounter{enumi}\theenumi\label{red.srs}] (Red series) We divide the whole period 1962--2024 into $4$ subintervals $[t_i, t_{i+1}]$ (about $15.75$ years each). For $t$ in  $[t_i, t_{i+1}]$, we let $\mathcal{I}_{t} = \mathcal{A}_{t_i}^K \cap \mathcal{A}_t$, i.e., the (still existing) stocks which were one of the largest $K$ on day $t_i$. Note that $|\mathcal{I}_{t}| < K$ if some of the stocks were delisted before day $t$. Interestingly, we see that {\it diversity generally decreases} in all subintervals. In other words, the capital distribution with respect to $\mathcal{I}_t$ tends to become more and more concentrated. Intuitively, this is due to the widely differing growth rates of the stocks and the eventual delisting of many of the securities. In each of the nearly 16 year periods, anywhere from 19\%--47\% of the securities that were originally the largest exit the market (corresponding to a delisting rate of 1--3\% per year).%
\item[\refstepcounter{enumi}\theenumi\label{grey.srs}] (Grey series) For each subinterval $[t_i, t_{i+1}]$ in \ref{red.srs}, we pick $K$ stocks $\mathcal{I}_{t_i}$ randomly in $\mathcal{A}_{t_i}^{M}$, where $M = 1000$, and, for $t \in [t_i, t_{i+1}]$, let $\mathcal{I}_t = \mathcal{I}_{t_i} \cap \mathcal{A}_t$. That is, we track the diversity of the randomly chosen stocks as a closed system. We generate $25$ batches for each subinterval. Again we see that the diversity of $\mathcal{I}_t$ generally decreases over time. Also, the randomly selected subuniverses are generally less diverse than $\mathcal{A}_{t_i}^K \cap \mathcal{A}_t$, which in turn is generally less diverse than $\mathcal{A}^K_t$.
\end{enumerate}
}

Since the capital distribution is ``stable'' and diversity is a function of the capital distribution, we may say, rather vaguely, that ``diversity is stable'' as a corollary to Stylized Fact \ref{fact:1}. Saying ``diversity is mean reverting'' requires the existence of a (possibly varying) equilibrium value which we fail to establish convincingly. On the other hand, we summarize the results of \ref{red.srs} and \ref{grey.srs} in the  stylized fact:

\begin{fact} \label{fact:3}
Diversity of a fixed collection of stocks tends to decrease over time.
\end{fact}

\begin{figure}
\includegraphics[scale = 0.8]{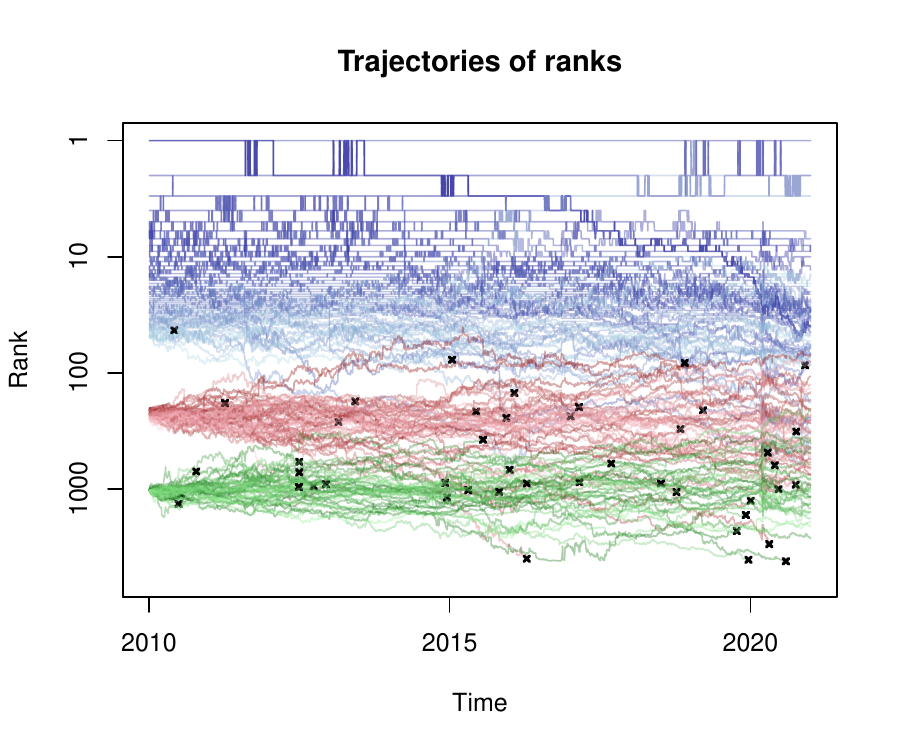}
\vspace{-0.3cm}
\caption{Ranks (in log-scale) of three groups of $50$ stocks (blue, red, green) over the period 2010--2020. Delisting events are marked with crosses.}
\label{fig:rank.trajectory}
\end{figure}

To better understand the results in \ref{red.srs} and \ref{grey.srs}, consider Figure \ref{fig:rank.trajectory} in which we track the ranks of three groups of $50$ stocks over the period 2010--2020. The initial ranks of the groups are respectively $1\leq k \leq 50$, $201\leq k \leq 250$ and $1001 \leq k \leq 1050$. We see that the top group, corresponding to the largest companies, remain mostly on top even after $10$ years. On the other hand, the ranks of the lower groups are much more volatile. For a fixed universe, this dispersion (as well as delistings) causes diversity to decrease over time, i.e., its capital distribution becomes more and more concentrated. More empirical properties related to ranks will be discussed in Section \ref{sec:rank.based.properties} where we identify an overall decreasing trend in the ranking of individual stocks and quantify the intensity of rank switching.

As a corollary of Stylized Fact \ref{fact:3}, which summarizes \ref{red.srs} and \ref{grey.srs} above, the apparent stability of the capital distribution, and hence diversity, is fueled by the constant entrances and exits of stocks over time. We emphasize that this observation is absent in the classical SPT model (which assumes a fixed collection of stocks) as well as  rank-based diffusion models such as \eqref{eqn:rank.based.diffusion}. We believe it is of substantial theoretical and practical interest to include entrances and exits in rank-based models. See Section \ref{sec:rank.transition.probabilities} for a further discussion.

Due to the effect of market diversity on the relative performance of actively managed portfolios (see Remark \ref{rmk:div.and.perf}), a natural question is to predict its future values. In \cite{AFF07}, the authors developed a generalized tree-structured model using macroeconomic information and reported outperformance in predictive power relative to standard time series methods. Even simple models may be quite useful. For example, the authors of \cite{TM21} implemented a regression model (motivated by the theoretical decomposition \eqref{eqn:diweighted.portfolio}) to devise a rule to decrease the drawdown of the equal-weighted portfolio by switching to a capitalization-weighted portfolio when diversity is expected to drop.\footnote{We remark that transaction costs are neglected in \cite{TM21}.} We note, however, that modelling diversity as a standalone time series is different from building a model for a system of stocks whose diversity exhibits realistic behaviours. Again, we stress the importance of economic models which explain why the capital distribution curve (and hence diversity) behaves in the way it does; they allow us to understand if the observed stability is truly a long term invariant or may shift subject to changes in the market (e.g.~the rise of massive tech companies and artificial intelligence). Another interesting problem, which may be of practical interest, is to use option prices (of stocks, indices and ETFs) to infer market participants' perceptions of future values of market diversity and possibly other quantities such as intrinsic volatility. Finally, we mention the problem of developing alternative measures of diversity which take into account additional features such as sectors and statistical similarities between different stocks,\footnote{This problem was suggested by Martin Larsson.} as well as systematic methodologies (including financial interpretations) for comparing the diversities of markets with different sizes.

\section{Intrinsic market volatility} \label{sec:excess.growth.rate}
The conventional or {\it absolute} aggregate volatility of an equity market usually refers to the volatility of its market index. For the US equity market, it is standard to use the past-looking historical volatility (e.g. standard deviation) of the returns of S\&P 500 and the forward-looking VIX (CBOE Volatility Index) which is computed using option prices. In contrast, by {\it intrinsic} market volatility (following the terminology in \cite{FK09}) we mean {\it relative volatility} among the stocks. For example, if all stocks go down by $20\%$ there is large absolute volatility but no relative volatility -- all equity portfolios yield the same return regardless of their weights. On the other hand, the market index may stay constant even if the stocks have widely differing returns. Thus relative volatility  is {\it necessary} for an actively managed portfolio to outperform the market,\footnote{See \cite{FKR18} for a theoretical study about its sufficiency.} and provides complementary information about market behaviours relevant to trading decisions. In this section, we quantify intrinsic market volatility using the concept of {\it excess growth rate} \cite{FS82}, which is also known in the finance literature as the {\it diversification return} \cite{BF92, Q18}.

\subsection{Excess growth rate} 
 We define the excess growth rate (EGR) using the formulation in \cite{PW13, PW16}. Consider a time interval $[t_0, t_1]$ and a collection $\mathcal{I}_{t_0} \subset \mathcal{A}_{t_0}$, such that stock $i \in \mathcal{I}_{t_0}$ has log-return $r_i(t_0, t_1) \in \mathbb{R}$ over $[t_0, t_1]$. Given a probability vector $w = (w_i)_{i \in \mathcal{I}_{t_0}} \in \mathcal{P}(\mathcal{I}_{t_0})$, we define the {\it excess growth rate} $\gamma_{w}(t_0, t_1)$ over $[t_0, t_1]$ and weighted by $w$, by
\begin{equation} \label{eqn:excess.growth.rate}
\gamma_w(t_0, t_1) := \log \Big( \sum_{i \in \mathcal{I}_{t_0}} w_i e^{r_i(t_0, t_1)} \Big) - \sum_{i \in \mathcal{I}_{t_0}} w_i r_i(t_0, t_1).
\end{equation}
Financially, $\gamma_w(t_0, t_1)$ is the difference between the log-return of the portfolio with weights $(w_i)$ and the weighted average log return of the underlying assets; this explains why it is called the {\it excess} growth rate. Jensen's inequality implies that $\gamma_w(t_0, t_1) \geq 0$ and (when $w_i > 0$ for all $i$) $\gamma_w(t_0, t_1) = 0$ if and only if all $r_i(t_0, t_1)$ are the same. Thus we may regard $\gamma_w(t_0, t_1)$ as a measure of {\it realized} relative volatility over the period $[t_0, t_1]$.\footnote{If $r_i(t_0, t_1) = -\infty$ and $w_i > 0$ then \eqref{eqn:excess.growth.rate} gives $\gamma_w(t_0, t_1) = \infty$. The excess growth rate never blows up to $\infty$ in our computation.} Common choices of $w$ include the capitalization weights $w_i = \mu_i^{\mathcal{I}_{t_0}}(t_0)$ and the equal weights $w_i = \frac{1}{n}$, where $n = |\mathcal{I}_{t_0}|$. If we are given a time grid $\{t_{\ell}\}_{{\ell} \geq 0}$, universes $\{\mathcal{I}_{t_{\ell}}\}_{\ell \geq 0}$ and weights $w(t_{\ell}) \in \mathcal{P}(\mathcal{I}_{t_{\ell}
})$, we define the {\it cumulative excess growth rate} $\Gamma_w(t)$  with respect to the given time grid and weights by $\Gamma_w(t_0) = 0$ and 
\begin{equation} \label{eqn:excess.growth.rate.aggregate}
\Gamma_w(t_{\ell}) := \Gamma_w(t_{\ell-1}) + \gamma_{w(t_{\ell-1})}(t_{\ell-1}, t_{\ell}), \quad \ell \geq 1.
\end{equation}
In a continuous time framework where stock prices are modelled by continuous semimartingales (see Appendix \ref{sec:appendix.spt}) the cumulative excess growth rate (as the partition size tends to zero) can be modelled in terms of a quadratic variation, see \eqref{eqn:dGamma} for the explicit limiting expression.

\begin{figure}
\includegraphics[scale = 0.6]{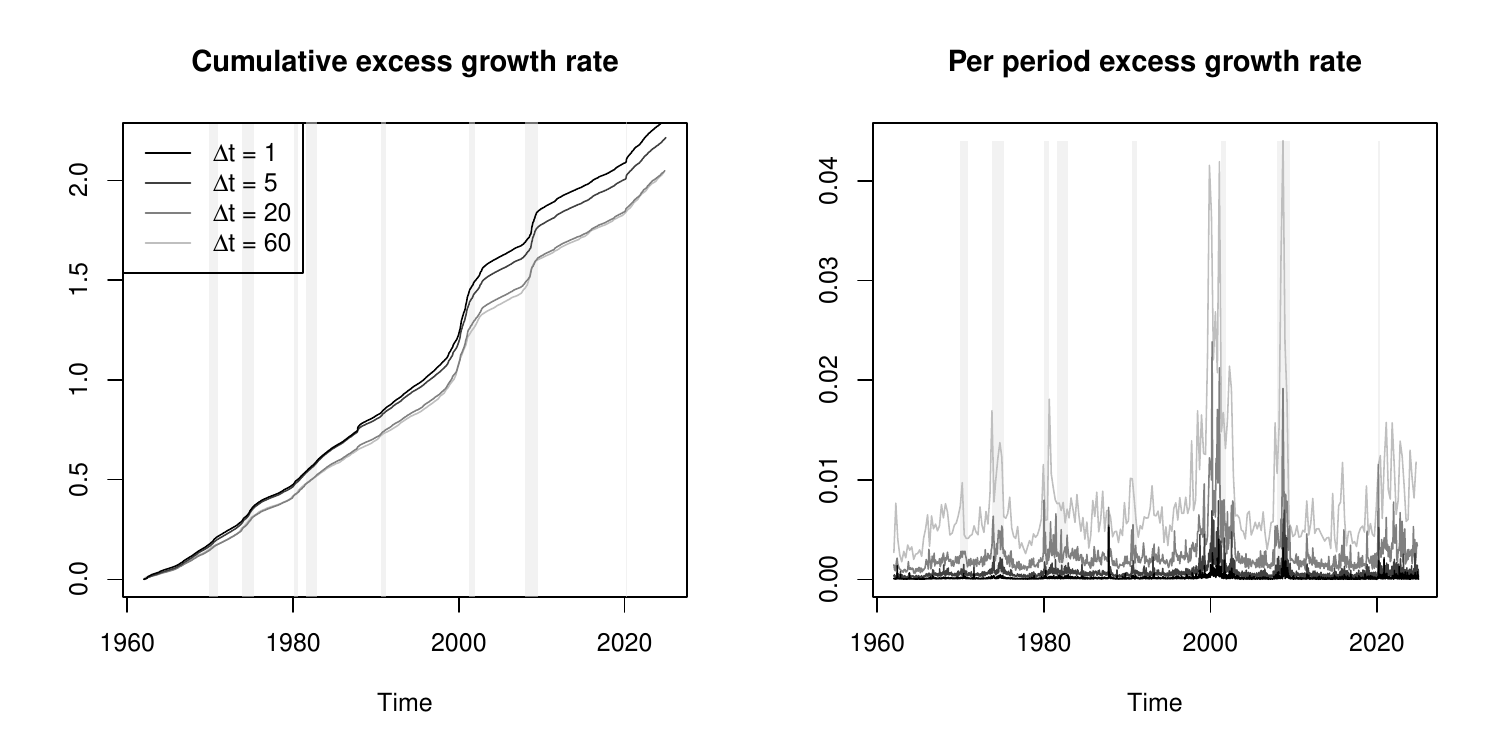}
\vspace{-0.3cm}
\caption{Left: Cumulative excess growth rate $\Gamma_w(t)$ for  $w(t) = \mu^{\mathcal{A}_t^{1000}}_t$, at various frequencies. Right panel: Per period excess grow rate $\gamma_w(t_{\ell}, t_{\ell + 1})$ for the same frequencies. The vertical shades indicate periods of recessions defined by the National Bureau of Economic Research.}
\label{fig:EGR}
\end{figure}

To further motivate the excess growth rate we discuss some of its properties:
\begin{itemize}
\item[(i)] The excess growth rate is {\it num\'{e}raire invariant}, i.e., $\gamma_w$ remains unchanged if the asset prices, and hence their returns, are measured with respect to another num\'{e}riare such as a benchmark portfolio \cite[Lemma 3.2]{PW13}. This makes the excess growth rate an appropriate measure of relative volatility.\footnote{Num\'{e}raire invariance uniquely characterizes the excess growth rate among a class of distance-like quantities called {\it $L$-divergences}; see \cite[Example 3.10]{PW18}.}
\item[(ii)] When $r_i(t_0, t_1) \approx 0$ for each $i$ the following quadratic Taylor approximation holds:
\begin{equation} \label{eqn:EGR.approximation}
\gamma_{w}(t_0, t_1) \approx \frac{1}{2} \left( \sum_{i \in \mathcal{I}_{t_0}} w_i r_i(t_0, t_1)^2 - \Big( \sum_{i \in \mathcal{I}_{t_0}} w_i r_i(t_0, t_1) \Big)^2 \right).
\end{equation}
Thus $\gamma_{w}(t_0, t_1)$ can be approximated by half of the {\it variance} of the log returns of the stocks when weighted by $w$. 
\item[(iii)] Fix a time grid $\{t_{\ell}\}$ with initial time $T_0$ and final time $T_1$, $\mathcal{I} \subset \mathcal{A}_{T_0}$, and a probability vector $w$ over $\mathcal{I}$. Summing \eqref{eqn:excess.growth.rate} over time gives the decomposition
\begin{equation} \label{eqn:discrete.decomposition}
\log \Big( \prod_{\ell} \sum_{i \in \mathcal{I}} w_i e^{r_i(t_{\ell}, t_{\ell + 1})} \Big)= \sum_{i \in \mathcal{I}} w_i r_i(T_0, T_1) + \Gamma_{w}(T_0, T_1),
\end{equation}
which is the discrete analogue of the continuous time portfolio value decomposition in \eqref{eqn:portfolio.basic.decomp}. The left hand side of \eqref{eqn:discrete.decomposition} is the log-return $r_w(T_0, T_1)$ of the portfolio rebalanced to the {\it constant} weight $w$ at each time $t_{\ell}$ (here and in (iii) we neglect transaction costs). Everything else equal, a large relative volatility is favourable to a rebalanced portfolio compared to a capitalization-weighted one. This observation underlies the idea of {\it volatility pumping} \cite{BNW15, dempster2007volatility, PW13, Q18} and can be formalized mathematically using the concept of {\it functional portfolio generation} in SPT \cite{F02, fernholz1999portfolio, PW16, W19}.\footnote{For a general portfolio (not constant weighted or functionally generated), pathwise decompositions such as \eqref{eqn:discrete.decomposition} do not apply and a direct attribution of performance to its excess growth rate is less clear. See  
\cite{chambers2014limitations, willenbrock2011diversification} and the references therein for discussions about the limitations of the excess growth rate or diversifcation return in explaining ``alpha''. In this section, we simply use the excess growth rate as a measure of intrinsic volatility. In Section \ref{sec:backtest} we relate the excess growth rate with the performance of the diversity-weighted portfolio which is functionally generated.} Thus it is natural to consider maximization of excess growth rate subject to suitable constraints \cite[Example 1.1.7]{F02}. For example, risk-adjusted performance of portfolios optimized with respect to excess growth rate was investigated in \cite{mantilla2022can}, and a generalized efficient frontier was constructed in \cite{ding2023optimization}.

\item[(iv)] The excess growth rate depends on the {\it sampling} or {\it rebalancing frequency} which may be non-constant over time. For example, given three time points $t_0 < t_1 < t_2$ and $w$, the excess growth rate $\gamma_{w}(t_{0}, t_2)$ is generally different from the sum $\gamma_{w}(t_{0}, t_1) + \gamma_{w}(t_{1}, t_2)$. When the former is larger it is better not to rebalance.\footnote{A geometric interpretation based on a {\it generalized Pythagorean theorem} is given in \cite{PW18}.} To explain this, suppose in the context of (ii) $\{\tilde{t}_{m}\}$ is another time grid over $[T_0, T_1]$. If $\tilde{r}_w(T_0, T_1)$ is the log return of the portfolio rebalanced to $w$ at times $\tilde{t}_m$, then 
\begin{equation} \label{eqn:discrete.decomposition2}
r_w(T_0, T_1) - \tilde{r}_w(T_0, T_1) = \Gamma_w(T_0, T_1) - \tilde{\Gamma}_w(T_0, T_1),
\end{equation}
where $\tilde{\Gamma}_w$ is the cumulated excess growth rate with respect to the time grid $\{\tilde{t}_{m}\}$. Thus the {\it spread} in \eqref{eqn:discrete.decomposition2} captures the effect of the frequency of rebalancing in the absence of transaction costs. The authors of \cite{F07} observed in a high-frequency setting that the cumulative excess growth rate increases as the frequency of rebalancing increases and devised a long-short hedging portfolio to exploit the spread.
\end{itemize}

\begin{figure}
\includegraphics[scale = 0.6]{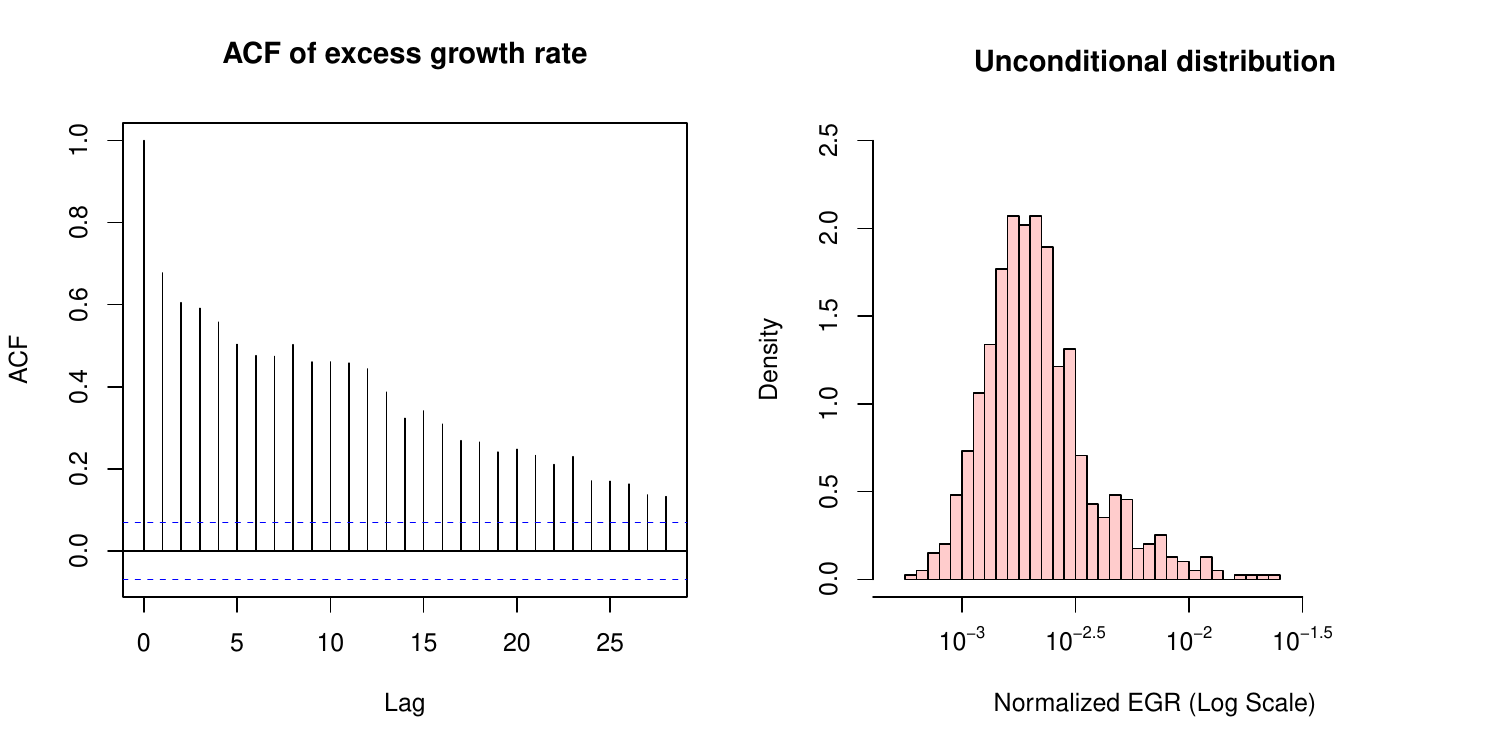}
\vspace{-0.3cm}
\caption{Empirical properties of $\gamma_w(t)$ in the context of Figure \ref{fig:EGR}, where $\Delta t = 20$.
Left: Sample autocorrelation function.  Right: Unconditional distribution of $\frac{1}{\Delta t} \gamma_w(t,t+\Delta t)$.}
\label{fig:EGR2}
\end{figure}

\subsection{Empirical results}
In Figure \ref{fig:EGR} we consider the per period excess growth rate $\gamma_w$ (right)  and the cumulative excess growth rate $\Gamma_w$ (left) at various frequencies, where $w(t) = \mu^{\mathcal{A}_t^K}(t)$ is the capitalization weights relative to the top $K = 1000$ stocks. We  consider the frequencies $\Delta t = 1$ (daily), $5$ ($\sim$weekly), $20$ ($\sim$monthly) and $60$ ($\sim$quarterly).\footnote{Here, and in the backtests in Section \ref{sec:backtest}, we use constant intervals in trading days instead of calendar days to mitigate possible calendar effects.} We observe that the cumulative excess growth rate grows roughly linearly over time (about $3.5\%$ per year when $\Delta t = 5$). Nevertheless, as seen in Figure \ref{fig:EGR} (right) and the sample autocorrelation function shown in Figure \ref{fig:EGR2} (left), the excess growth rate exhibits clear ARCH effects\footnote{By an ``ARCH effect'' we mean that the time series has positive serial correlation at multiple scales.} at all frequencies considered, and is large during the financial crises in 2000 and 2008 among other significant market events. We stress that stochastic volatility of this sort is absent in all rank-based diffusion models  studied so far, hence it is a promising direction to extend these models to capture the time series properties of relative volatility (alongside stability of the capital distribution curve).

\begin{fact} \label{fact:4}
Market excess growth rate is correlated with absolute market volatility and clusters in time.
\end{fact}

The unconditional (marginal) distribution of the excess growth rate is highly skewed towards the right. In Figure \ref{fig:EGR2}(right) we plot the empirical distribution of the normalized excess growth rate $\frac{1}{\Delta t} \gamma_w(t,t+\Delta t)$ for $\Delta t=20$ (the distributions for other frequencies are similar). The sample skewness of $\gamma_w(t,t+\Delta t)$ is $4.3$ and the excess kurtosis is $27.3$. There is moderate positive correlation between excess growth rate (relative volatility) and market (absolute) volatility. For example, for $\Delta t = 5$ the empirical correlation between $\gamma_{w}$ and the squared capitalization-weighted return is about $0.41$; the Winsorized version, which corrects for outliers, is about $0.24$.\footnote{We use here the function \texttt{corHuber()} from the R package \texttt{robustHD}.}

\begin{figure}
\includegraphics[scale = 0.6]{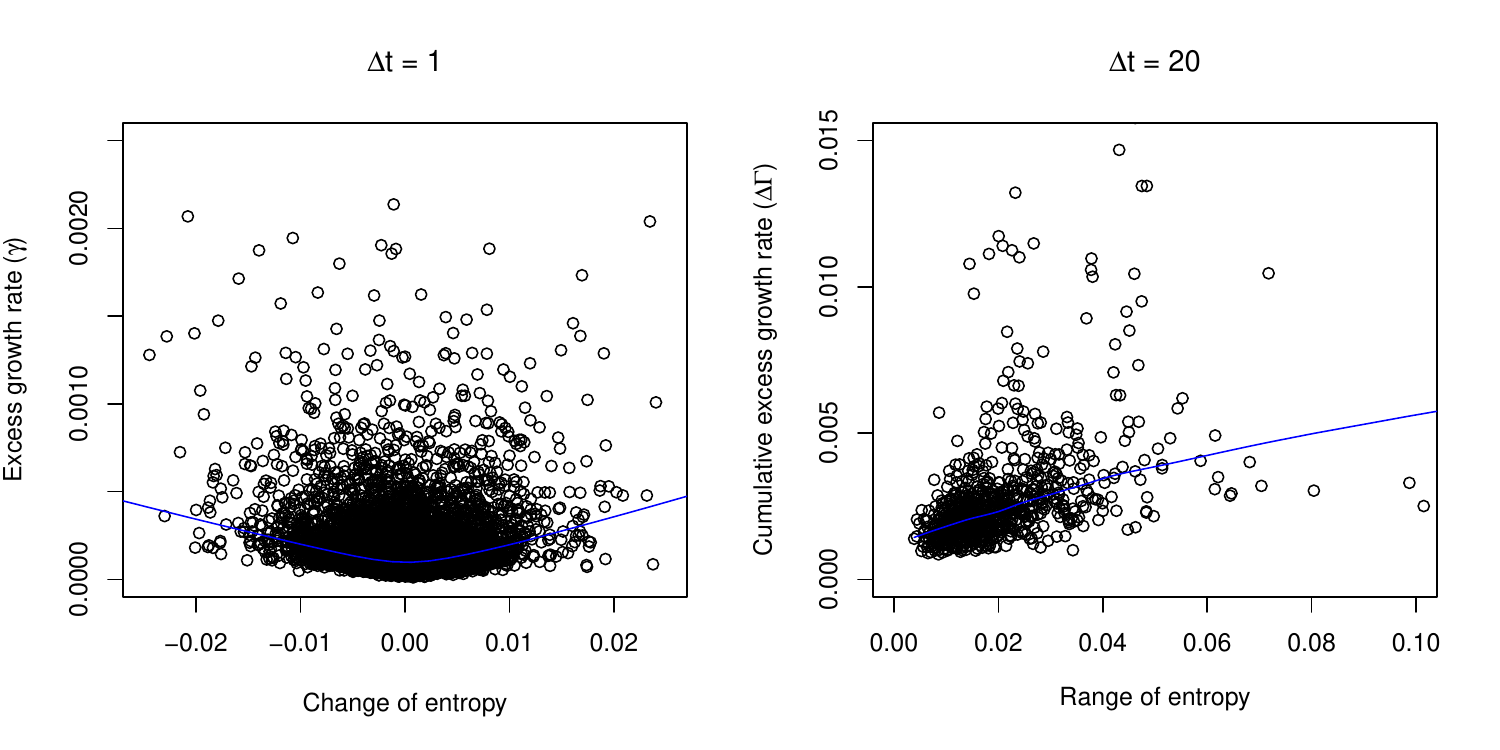}
\vspace{-0.3cm}
\caption{Joint behaviours of market diversity and intrinsic volatility of the largest $K = 1000$ stocks (varying over time). Left: Scatter plot of daily excess growth rate versus change of log entropy. Right: Scatter plot of cumulative (daily) excess growth rate versus range ($\max - \min$) of entropy over the same period; each point covers $20$ trading days. The blue curves are fitted with locally weighted scatterplot smoothing (LOWESS). In both plots there are several outliers outside the ranges shown.} 
\label{fig:joint}
\end{figure}

\medskip

Next we turn to joint behaviours of market diversity (here quantified by market entropy) and intrinsic volatility (excess growth rate). Recall that intrinsic volatility is necessary for a rebalanced portfolio (such as the diversity-weighted portfolio simulated in Section \ref{sec:backtest}) to outperform a capitalization-weighted benchmark. However, such a portfolio is exposed to risks posed by changes in market diversity. In Figure \ref{fig:joint} (left) we plot the daily excess growth rate $\gamma_w$ (where $w$ is the capitalization weight vector) and the daily change in diversity, again for the top $1000$ stocks on each day. The underlying pattern is not visually obvious due to the large number of data points. A local regression (LOWESS, see e.g.~\cite[Chapter 1]{efron2021computer}) reveals a systematic dependence which is roughly symmetric in the sign of the change in diversity. The relationship is more evident if we zoom out in time. In Figure \ref{fig:joint} (right) we consider instead time intervals with length $\Delta t = 20$ trading days. For each interval $[t_{\ell}, t_{\ell+1}]$, we plot the cumulative daily excess growth rate $\Delta \Gamma_w = \Gamma_w(t_{\ell+1}) - \Gamma_w(t_{\ell})$ versus the range $\max_{t \in [t_{\ell}, t_{\ell+1}]} {\bf H}(\mu(t)) - \min_{t \in [t_{\ell}, t_{\ell+1}]} {\bf H}(\mu(t)) \geq 0$ of diversity. Again we observe a positive dependence. Thus we state the next stylized fact as follows:

\begin{fact} \label{fact:5}
Market excess growth rate tends to be larger when market diversity is volatile and vice versa.
\end{fact}

\section{Rank-based properties} \label{sec:rank.based.properties}
In this section we study how the behaviours of stocks depend systematically on their relative ranks with respect to market capitalization.\footnote{This is analogous to the practice of forming {\it deciles} in empirical finance: for each period, group the assets after ordering them with respect to some criterion. Here the criterion is rank (by market capitalization) and we will examine quantities indexed by rank.} In stochastic portfolio theory there are two main motivations for the study of rank-based properties. First, as mentioned in Section \ref{sec:capdist.and.diversity}, ``projecting'' to the space of ranked market capitalization or weights (thus neglecting name-based features as in the rank-based diffusion system \eqref{eqn:rank.based.diffusion}) allows for more parsimonious modelling of some features of market capitalizations and the capital distribution curve. Second, it is common to restrict the investment universe to a top segment of stocks by market capitalization (one reason is that these stocks are more representative of the market and are also  more liquid). When strictly enforced, this means selling a stock when it drops out of the universe and buying a stock when it enters. Thus part of the turnover of the portfolio is directly related to the ``intensity'' of rank switching near the cut off rank.\footnote{Rank switching also plays an important role when the benchmark is a capitalization-weighted portfolio consisting of the largest $K$ stocks (rather than the entire market). {\it Leakage} refers to the loss in wealth  due to renewing the portfolio's constituent stocks. We refer the reader to \cite[Example 4.3.5]{F02} and \cite{X20} for a detailed discussion of leakage and its computation.
}

\subsection{Rank-based volatility} \label{sec:rank.volatility}

\begin{figure}
\includegraphics[scale = 0.6]{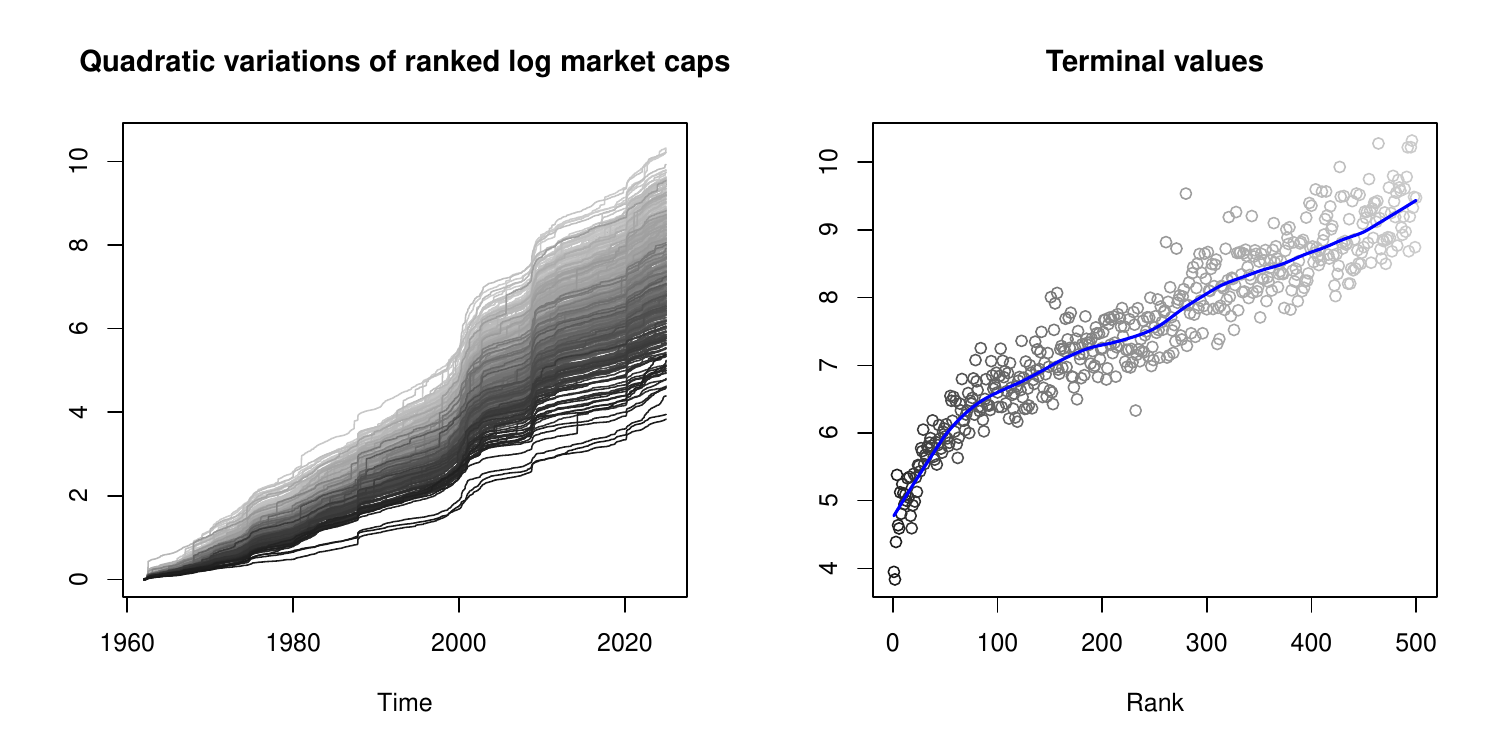}
\vspace{-0.3cm}
\caption{Left: Discrete quadratic variations $QV_k(t)$, defined by \eqref{eqn:QV}, for $k = 1$ (dark grey) to $k = 500$ (light grey). Right: The terminal values $QV_k(t_{\text{term}})$. Smoothed values are shown by the blue curve.}
\label{fig:QV}
\end{figure}

We first provide a quantitative illustration that smaller stocks are on average more volatile. Recall that $X_i(t)$ is the market capitalization of stock $i$ on day $t$. For each rank $k \in \{1, \ldots, 500\}$ we consider a {\it discrete quadratic variation} $QV_k(t)$ of the log market capitalization at rank $k$. On day $t$, suppose stock $i_k(t)$ has rank $k$. We define
\begin{equation} \label{eqn:QV}
QV_k(t + 1) = QV_k(t) + (\log X_{i_k(t)}(t+1) - \log X_{i_k(t)}(t))^2,
\end{equation}
and if stock $i_k(t)$ is not traded on day $t + 1$ (e.g. delisted) we simply let $QV_k(t+1) = QV_k(t)$. We initialize $QV_k(t_{\text{init}}) = 0$ where $t_{\text{init}}$ is the first trading day (1962-01-02) of the data-set. Since $i_k(t)$ varies over time, multiple stocks contribute to $QV_k(t)$ for each rank $k$. We show the results in Figure \ref{fig:QV}, in which the left panel plots the time series and the right panel plots the terminal values. We observe, as expected, that smaller stocks are on average more volatile and the effect is non-trivial: at the terminal time $t_{\text{term}}$ we have $QV_{500}(t_{\text{term}}) \approx 9.48$ and $QV_1(t_{\text{term}}) \approx 3.95$. These quantities correspond to an average annual quadratic variation of $0.15$ and $0.06$, respectively.

\begin{fact} \label{fact:6}
Smaller stocks are on average more volatile.
\end{fact}

In \cite[Section 5]{CW22b}, the first and last author analyzed further the empirical {\it distributions} of returns arranged by ranks. More precisely, if $R_i(t)$ is the return of stock $i$ on day $t$, we define for each rank $k$ an empirical return distribution $\nu_k$ by $\nu_k = \frac{1}{T} \sum_t \delta_{R_{i_k(t)}(t)}$, where $T$ is the number of observations and $\delta_{x}$ is the point mass at $x$. We call $(\nu_k)_k$ a {\it distributional data-set} since each data point $\nu_k$ is itself a probability distribution (on $\mathbb{R}$). It is natural to study how the distribution $\nu_k$ varies in $k$. Performing geodesic principal component analysis using the {\it Wasserstein geometry} in optimal transport \cite{BGKL17}, we showed that smaller stocks are not only more volatile but their returns are also more positively skewed; these effects are captured by the first and second geodesic components. On the other hand, there is a lack of evidence that the (log) growth rates of stocks have a material systematic dependence on their ranks. See, for example \cite[Figure 5.4]{F02}, \cite[Figure 1]{BFPRS19}, \cite[Figure 1]{FF20} and \cite[Figure 3]{CW22b}. In the context of the rank-based model \eqref{eqn:rank.based.diffusion}, this means that the rank-based growth rates do not have a clear dependence on $k$. Here, we remark that the difficulty of estimating growth rates accurately is a well-known challenge in financial econometrics.

\subsection{Rank transition probabilities} \label{sec:rank.transition.probabilities}
In Figure \ref{fig:rank.trajectory} we visualize how the ranks of a collection of stocks change over time. Now we attempt a more quantitative description by estimating empirically the {\it transition probabilities} of daily rank transitions.

\begin{figure}
\includegraphics[scale = 0.6]{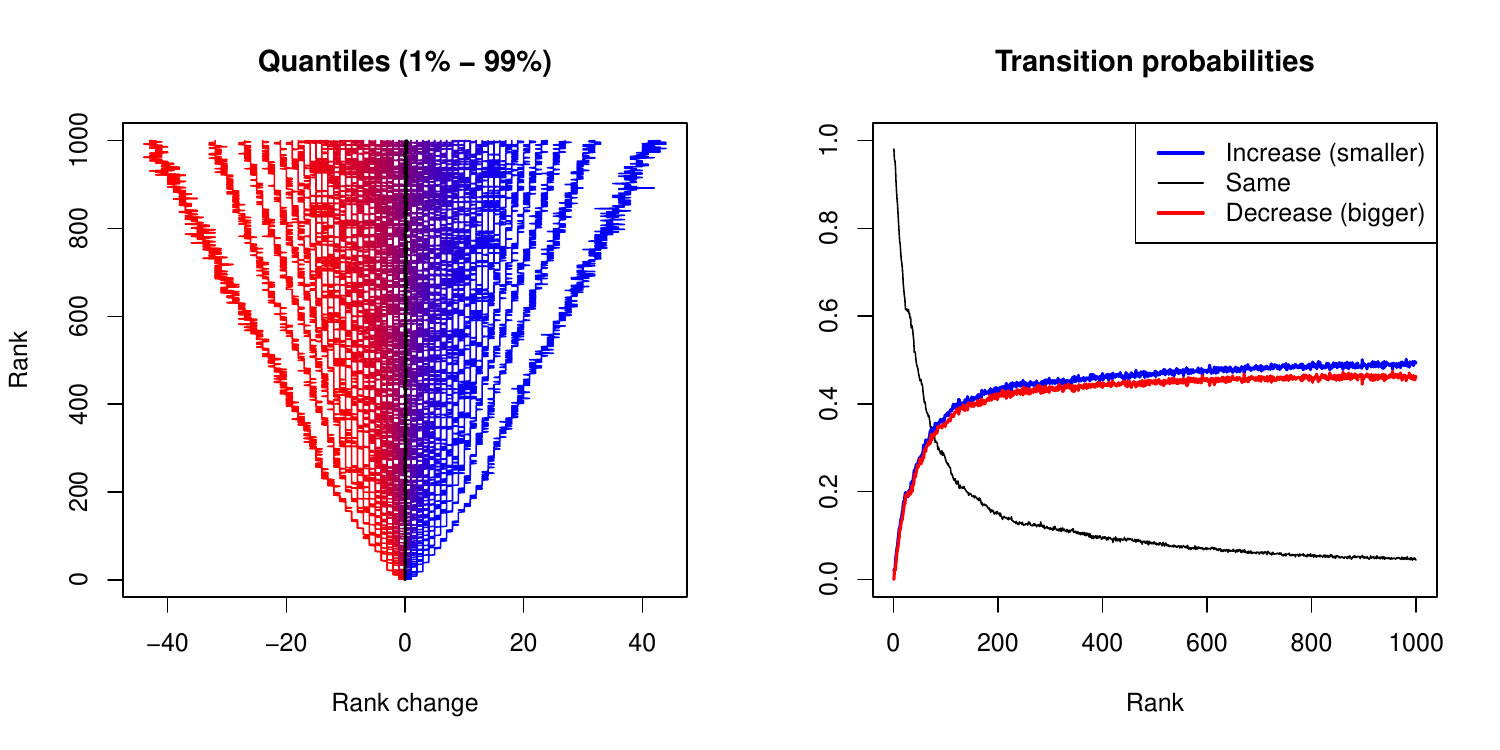}
\vspace{-0.3cm}
\caption{Left: Empirical distributions of daily rank changes. Shown here are the quantiles (at probabilities $0.01, 0.02, \ldots, 0.99$, from blue to red) of rank change (truncated to $[-50, 50]$) at each rank from $1$ to $1000$. The black curve gives the mean for each rank (smoothed values range from $0.024$ at rank $1$ to $0.156$ at rank $1000$). Right: Empirical probabilities of positive (blue), null (black) and negative (red) daily rank changes as a function of rank.}
\label{fig:transitions}
\end{figure}

We do this straightforwardly. For each day $t$ (over the entire sample period) and rank $k \in \{1, \ldots, 1000\}$, we find the stock $i_k(t)$ at rank $k$. If its rank is $\ell$ on day $t + 1$, the {\it rank change} is $\ell - k$. According to our convention, a positive change in rank means that the stock becomes {\it relatively} smaller. For example, for $k = 2$ the possible values of rank change are $-1, 0, 1, 2, \ldots$. In the case of delisting the rank change is $+\infty$. With this we obtain a matrix of relative frequencies of rank changes.\footnote{For computational purposes we consider rank changes from $-50$ to $50$ and truncate all values beyond at the boundary values. Thus delisting means $50$ in Figure \ref{fig:transitions}(left) and ``Increase'' in Figure \ref{fig:transitions}(right)%
} We emphasize that the rank changes estimated this way are affected by entrances and exits. For example, if a new stock enters at rank $11$, the ranks of all smaller stocks will increase by $1$ if there are no other changes.

We show the results in Figure \ref{fig:transitions}. As expected, we observe that rank changes are statistically larger in magnitude as the rank increases. At rank $1000$ the average daily rank change (after truncation and smoothing) is about $0.156$; the value at rank $1$ is about $0.024$. These values depend on the handling of delisting. The pattern is clearer if we consider simply the {\it sign} (positive, zero or negative) of daily rank change, resulting in empirical probabilities $p_{+}(k)$, $p_{0}(k)$ and $p_{-}(k)$ for each rank $k$. In Figure \ref{fig:transitions}(right) we observe that $p_{+}(k) > p_{-}(k)$ (with a few exceptions) and the difference $p_+(k) - p_-(k)$ increases roughly linearly as $k$ increases. At rank $1000$, the difference is (after smoothing) about $0.029$. With this we state:

\begin{fact} \label{fact:7}
There is a tendency for a stock's capitalization rank to increase (i.e., become relatively smaller).
\end{fact}

At first sight, this tendency may appear to be inconsistent with the overall stability of the market and the lack of evidence that the growth rates of stocks depend systematically on their ranks. Again, we emphasize that the rank change includes the effects of entrances and exits visualized in Figure \ref{fig:Weights_Entrants_Exits_TS}. The patterns of the transition probabilities we observe in Figure \ref{fig:transitions} reflect a delicate balance between ``pure'' rank transitions, as well as entrances and exits. Our findings highlight the importance of entrances and exits in maintaining the stability of the equity market, and suggest the possibility of incorporating their effects in the formulation and calibration of rank-based models.

\begin{remark}
A finer analysis that estimates $p_+(k) - p_-(k)$ on separate time intervals reveals that the magnitude of $p_+(k) - p_-(k)$ (which is still generally positive) has been decreasing in recent years. This may be in part due to the general decrease in the size of the CRSP universe since the late 1990s. It will be interesting to see if this trend continues.
\end{remark}

\subsection{Intensity of rank switching} \label{sec:local.time.empirical}
The larger volatility of smaller stocks suggests that switching of ranks occurs more frequently among smaller stocks. To quantify this effect, we introduce an empirical quantity $\Lambda_k(t)$ which measures the (cumulative) intensity of rank switching at rank $k$. Its definition is motivated by a discretization of the {\it semimartingale local time} in stochastic calculus which we recall in Appendix \ref{sec:details}.\footnote{Although $\Lambda_k$ is motivated by the semimartingale local time, for the purposes of this paper we do not treat it as a statistical estimator (this assumes implicitly that the market is modeled by a continuous semimartingale).} We initialize each $\Lambda_k$ to start at $0$.

Let $Y_i(t) = \log X_i(t)$ be the log market capitalization of stock $i \in \mathcal{A}_t$ on day $t$, and let $Y_{(1)}(t) \geq Y_{(2)}(t) \geq \cdots$ be the decreasing order statistics. Also let $i_k(t)$ be the stock which has rank $k$ on day $t$. First consider the dynamics of $Y_{(1)}$. By definition, we have
\[
Y_{(1)}(t+1) - Y_{(1)}(t) = Y_{(1)}(t+1) - Y_{i_1(t)}(t) \geq Y_{i_1(t)}(t+1) - Y_{i_1(t)}(t).
\]
Given $\Lambda_1(t)$, we define $\Lambda_1(t+1) \geq \Lambda_1(t)$ by the identity
\begin{equation}  \label{eqn:discrete.local.time}
Y_{(1)}(t+1) - Y_{(1)}(t) = Y_{i_1(t)}(t+1) - Y_{i_1(t)}(t) + \frac{1}{2}(\Lambda_1(t+1) - \Lambda_1(t)).
\end{equation}
Note that $\Lambda_1(t+1) = \Lambda_1(t)$ if $Y_{(1)}(t+1) = Y_{i_1(t)}(t+1)$, i.e., stock $i_k(t)$ remains the largest stock on day $t + 1$. Otherwise, $\Lambda_1(t+1) - \Lambda_1(t) \geq 0$ measures the {\it amount of crossing}. Inductively, for each rank $k = 2, 3, \ldots$, we define $\Lambda_k(t+1)$ such that
\begin{equation} \label{eqn:discrete.local.time2}
\begin{split}
Y_{(k)}(t+1) - Y_{(k)}(t) &= (Y_{i_k(t)}(t+1) - Y_{i_k(t)}(t)) + \\
&\quad \frac{1}{2}\left( (\Lambda_k(t+1) - \Lambda_k(t)) - (\Lambda_{k-1}(t+1) - \Lambda_{k-1}(t)) \right).
\end{split}
\end{equation}
(This corresponds to \eqref{eqn:ranked.process.SDE} in the continuous time framework.) It is not difficult to verify that the solution to the system \eqref{eqn:discrete.local.time2} is given by  \[\Lambda_k (t+1)-\Lambda_k(t)= 2\sum_{j\leq k} Y_{(j)}(t+1) - 2\sum_{j\leq k} Y_{i_{j}(t)}(t+1)\geq0.\]
The positivity of the solution is obvious since the total mass of the largest $k$ elements at time $t+1$ must be bigger than any arbitrary collection of size $k$. We see that if $ \Lambda_k(t+1)-\Lambda_k(t)>0$ it means that the capitalization of the top $k$ stocks at time $t+1$ is strictly larger than the total capitalization of the stocks that used to occupy the top $k$ ranks at $t$ -- i.e., there is ``displacement from below" by stocks that used to occupy the lower ranks. We call $\Lambda_k$ defined by \eqref{eqn:discrete.local.time} and \eqref{eqn:discrete.local.time2} the {\it intensity of rank switching} at rank $k$.%

\begin{figure}
\includegraphics[scale = 0.6]{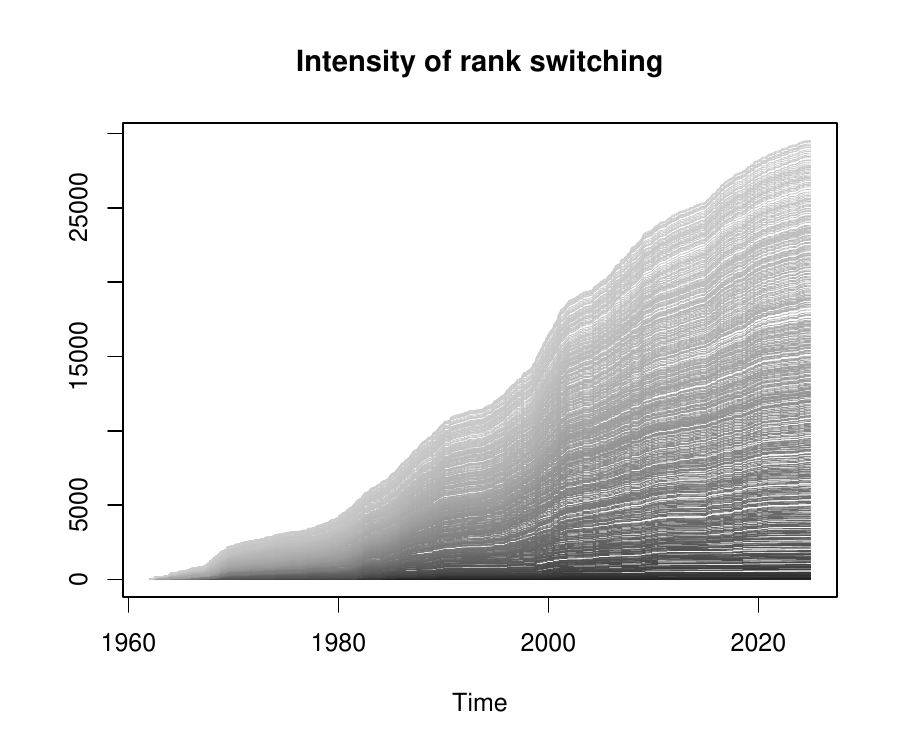}
\vspace{-0.3cm}
\caption{Time series of the intensity of rank switching $\Lambda_k$, from $k = 1$ (dark grey) to $k = 499$ (light grey).}
\label{fig:local.time}
\end{figure}

In Figure \ref{fig:local.time} we plot the intensity $\Lambda_k(t)$ as a function of time, for $k = 1, \ldots, 499$. We observe $\Lambda_k(t)$ increases roughly linearly in time and in $k$. Moreover, the intensity of rank switching is very small for the largest stocks ($k$ small). As seen in Figure \ref{fig:rank.trajectory}, the largest stocks mostly remain on top and rank switches only occur occasionally. These findings are consistent with the results (concerning the period $1990$--$1999$) presented in \cite[Section 5.2]{F02}. 

\begin{fact} \label{fact:8}
Rank switching occurs more intensely for small stocks.
\end{fact}

To end this section, we note that the authors of \cite{IL23} constructed a rank-based volatility stabilized model and calibrated it to fit empirical behaviours of rank-based volatility, intensity of switching and some other quantities. They were also able to characterize explicitly the resulting growth optimal portfolio in their model. Further improvements of this and related models will be helpful in optimizing and backtesting trading strategies for large portfolios. Additionally, the calibration of these models often relies on \textit{finite sample} and \textit{discrete time} estimators for the rank-based statistics of continuous time processes (e.g.~one can consider $\Lambda_k$ as an estimator for its continuous time counterpart in \eqref{eqn:ranked.process.SDE}). It is an open problem to study statistical properties of estimators of local times and their applications in the calibration of ranked particle systems.

\section{Performance of portfolios} \label{sec:backtest}
In this section, we show in a well-specified empirical setting that market diversity and intrinsic volatility, which are macroscopic quantities studied in the previous sections, are useful for explaining
the relative performance of certain portfolios (namely, diversity-weighted portfolios with various parameters) with respect to the capitalization-weighted portfolio. Our experiment is motivated by, and in some aspects extends, the empirical study in \cite{RX20}. In Section \ref{sec:portfolio.optimization.literature} we survey the literature on portfolio optimization in SPT.

\subsection{Set-up of the experiment}
The results of any backtesting experiment depend on many factors including market conditions during the sample period and the exact implementation of the portfolios. Here, our objective is to illustrate the effects of market diversity and intrinsic volatility in the most straightforward and hopefully convincing way.\footnote{The codes of our implementation are available at our Github repository and the reader is invited to vary the settings.} In stochastic portfolio theory, these quantities arise explicitly in decompositions of portfolio performance relative to the market (see \eqref{eqn:diweighted.portfolio} for the case of the diversity-weighted portfolio).  As will be explained below, our implementation includes proportional transaction costs and handles faithfully default and delisting events. In the following, by the {\it support} of a portfolio at time $t$ we mean the set $\mathcal{S}_t \subset \mathcal{A}_t$ of stocks held by the portfolio. At any given point in time our portfolios will hold at most $K = 500$ stocks.

\begin{itemize}
\item[(i)] We divide the entire sample period (1962--2024) into $63$ periods corresponding to the calendar years.\footnote{This frequency is chosen so that we may illustrate performance in a variety of market conditions. Moreover, the top segment $\mathcal{A}^K_t$ does not change drastically over $[t_{\ell}, t_{\ell+ 1}]$.} Denote by $t_\ell$ the beginning of the $\ell$th year, $\ell=1,\dots,63$. The experiment is repeated independently for each {\it window} $[t_{\ell}, t_{\ell+1}]$, in the sense that for $\ell$ fixed, all portfolios are initiated at time $t_{\ell}$ at the same value and trade until time $t_{\ell+1}$. We refrain from maintaining portfolios over long periods since this requires specific conventions and algorithms for updating the support of the portfolio.\footnote{Development of systematic methods is an interesting direction but is beyond the scope of this paper.}
In the descriptions below a window $[t_{\ell}, t_{\ell + 1}]$ is fixed.   
\item[(ii)] 
Our main {\it benchmark} portfolios are the market index tracking portfolios updated at some frequency, $f$. At time $t_{\ell}$, we initialize the portfolios with the capitalization weights $w_i(t_{\ell}) = \mu_i^{\mathcal{A}^K_{t_{\ell}}}(t_{\ell})$. Every $f$ trading days these portfolios rebalance to hold the market weights of the top $K=500$ stocks. If a stock delists in between rebalancing times, its delisting return (if provided) is treated as the last non-zero return on the security and the holding in this stock is not redistributed to other securities in the portfolio until the next rebalancing time. These portfolios are expected to behave similarly to the S\&P 500, a standard benchmark portfolio for the US market, when $[t_{\ell}, t_{\ell + 1}]$ is reasonably short.
\item[(iii)] We backtest {\it diversity-weighted portfolios} $w^{p, f}$ (see also \ref{eqn:diversity.weighted.portfolio}) parameterized by $p \in [0, 1]$ and a rebalancing frequency $f \in \{1, 2, \ldots\} \cup \{\infty\}$. It is initialized at $t_{\ell}$ to have weights 
\[
w_i^{p, f}(t_{\ell}) = \frac{ \Big( \mu_i^{\mathcal{A}^K_{t_{\ell}}}(t_{\ell}) \Big)^{p} }{ \sum_{j \in \mathcal{A}_{t_{\ell}}^K} \Big( \mu_j^{\mathcal{A}^K_{t_{\ell}}}(t_{\ell}) \Big)^{p} }, \quad i \in \mathcal{A}^K_{t_{\ell}}.
\]
The diversity-weighted portfolio, introduced in \cite{fernholz1999portfolio, FGH98}, is a key example of a {\it functionally generated portfolio} in SPT, where the portfolio weights are deterministic functions of the market weights. Note that the portfolio is equal-weighted\footnote{An equal-weighted portfolio may be used as another benchmark. For example, the S\&P 500 Equal Weight Index is the equal-weighted version of the usual S\&P 500. In \cite{platen2012approximating}, the author argued that under suitable conditions, the equal-weighted portfolio approximates the num{\'e}raire portfolio as the number of securities tends to infinity.} when $p = 0$ and capitalization-weighted when $p = 1$, so we may regard $p$ as an interpolation parameter. These portfolios are frequently used in empirical studies, see e.g. \cite{malladi2017equal, plyakha2012does, SV16, ZZR20}. The portfolio then rebalances and reinvests dividends (and cash proceeds from delistings) every $f$ trading days to the weights
\[
w_i^{p, f}(t) = \frac{ \Big( \mu_i^{\mathcal{S}_t}(t) \Big)^{p} }{ \sum_{j \in \mathcal{S}_t } \Big( \mu_j^{\mathcal{S}_t}(t) \Big)^{p} }, \quad i \in \mathcal{S}_t := \mathcal{A}_t^K, %
\]
while paying proportional transaction costs. Note that by construction $w^{1, f}$ are the benchmarks in (ii). In Section \ref{sec:backtest.results} we will report results for $p \in \{0, 0.01, \ldots, 1\}$ and  $f \in \{1, 2, 5, 10, \allowbreak 25, 50, 125, \infty\}$. We let $Z_{p, f}(t)$ be the wealth of the portfolio $w^{p, f}$, normalized to $1000$ at time $t_{\ell}$.
\item[(iv)] In (ii) and (iii), all trading activities (except initialization) incur {\it proportional transaction costs} following exactly the procedure described in \cite[Section 2.1]{RX20}. In view of \cite{CW22b, RX20}, we choose a cost of $ 0.25\%$ for all stocks on purchases and sales. The empirical study \cite{anand2013market} and the NASDAQ report \cite{tcost_nasdaq} also support a choice of cost on this order of magnitude.
\end{itemize}

\begin{remark}
Using our backtesting engine available on our repository, it is straightforward to simulate other portfolios such as the {\it entropy-weighted portfolio} \cite[Chapter 2]{F02} or the {\it additively generated portfolios} in introduced in \cite{karatzas2017trading}. We expect that our qualitative findings extend to these and other functionally generated portfolios that are long-only and overweight smaller stocks.
\end{remark}

\begin{figure}[t!]
    \centering

    \begin{subfigure}[c]{0.59\textwidth}
        \includegraphics[width=\textwidth]{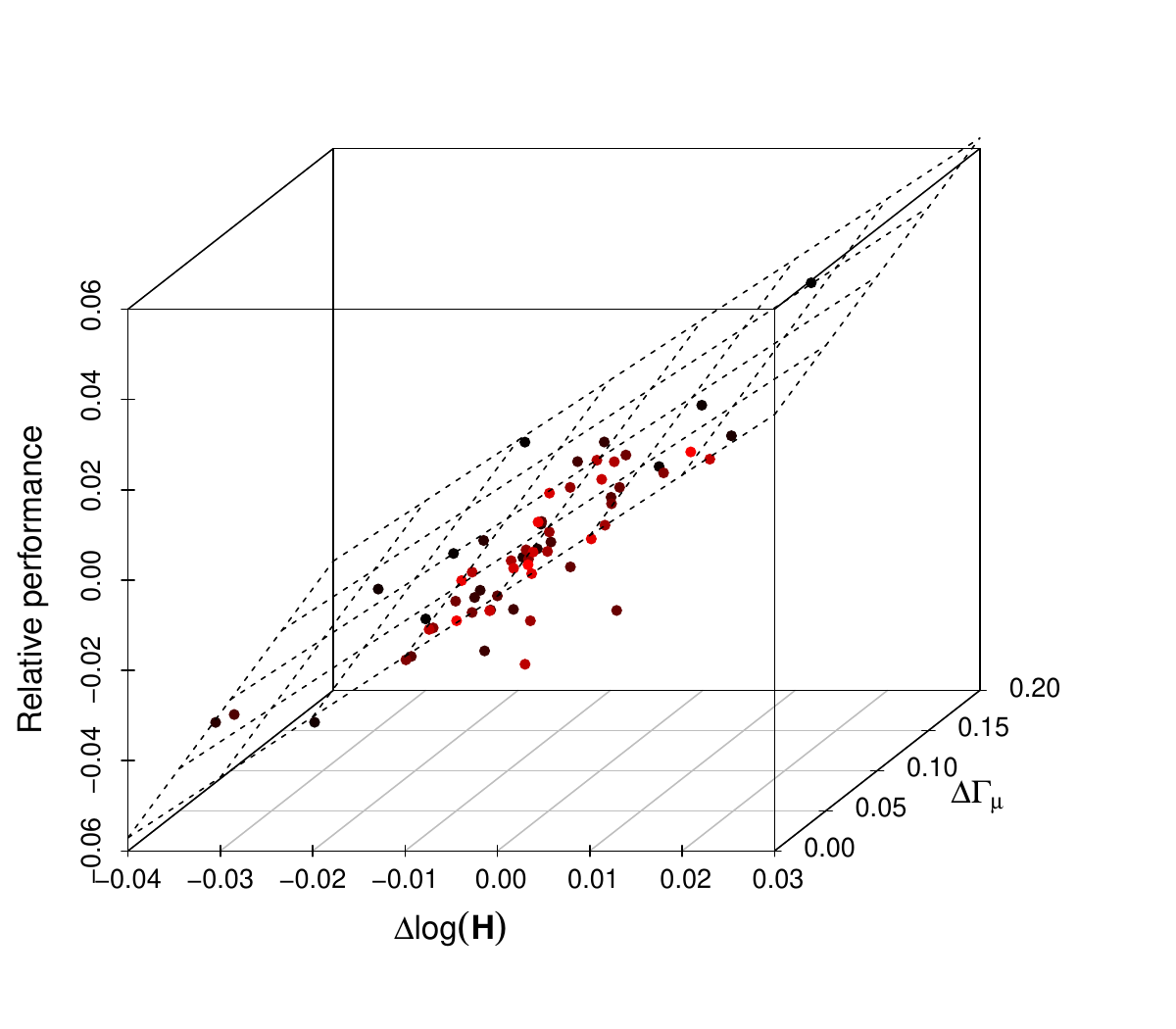}
        \label{fig:regression_plot}
    \end{subfigure}
    \begin{subfigure}[c]{0.4\textwidth}
        \begin{tabular}{c|c} %
        \\[-1.8ex]%
         & Coeff.\ ($\hat{\beta}_i$) \ Std.\ Error\\
        \hline \\[-1.8ex] 
        Intercept & $-0.004 \quad \quad (0.002)$  \\ 
        $\Delta\log(\mathbf{H})$ & $\ \ 1.339 \quad \quad  (0.092) $ \\ 
        $\Delta \Gamma_\mu$ & $\ \ 0.129 \quad \quad (0.044)$
        \end{tabular} 
        \label{fig:regression_table}
    \end{subfigure}
    \vspace{-0.5cm}
    \caption{Illustration of the linear regression \eqref{eqn:reg.model} concerning the annual relative performance of the equal-weight portfolio to the market index tracking portfolio when $f=10$ days, $\Delta \log (Z_{0,10}/Z_{1,10})$. Left: Scatter plot of the data with the fitted surface. Right: Regression coefficients $\hat{\beta}_i$ and standard errors (from ordinary least squares). }
    \label{fig:combined.reg}
\end{figure}

\subsection{Empirical results} \label{sec:backtest.results}

\subsubsection{Effect of diversity and intrinsic volatility in relative performance}
The above set-up yields, for each window $[t_{\ell}, t_{\ell+1}]$, a wealth process for each portfolio $w^{p, f}$, $p \in [0, 1]$ and $f \in \{1, 2, \ldots\}\cup \{\infty\}$. To underscore the relevance of diversity and intrinsic volatility in explaining the relative performance of portfolios, we regress the annual performance of the equal weight portfolio $w^{0, 10}$ relative to the captialization-weighted portfolio $w^{1, 10}$ (both rebalanced every $10$ days) against the annual changes in entropy and the (cumulative) {\it daily} market excess growth rate (of $\mathcal{A}^{K}_t$ for both quantities). Explicitly, we fit by ordinary least squares the linear model
\begin{equation}\label{eqn:reg.model}
\Delta \log(Z_{0,10}(t_\ell)/Z_{1,10}(t_\ell))\sim \beta_0 + \beta_1 \Delta \log\mathbf{H}(\mu_i^{\mathcal{A}_{t_\ell}^K}(t_\ell))+\beta_2 \Delta \Gamma_\mu(t_\ell),
\end{equation}
where $\Delta$ is the forward difference operator, $\Delta Z(t_\ell) = Z(t_{\ell+1})-Z(t_\ell)$. We illustrate the fit in Figure \ref{fig:combined.reg} (left) and report the estimated coefficients in Figure \ref{fig:combined.reg} (right). The $R^2$ and Adjusted $R^2$ values for the calibrated model are $0.78$ and $0.77$, respectively.

Under an idealized theoretical set-up in stochastic portfolio theory (see Appendix \ref{sec:appendix.spt}), a portfolio relative value decomposition reminiscent of the model in \eqref{eqn:reg.model} should hold.\footnote{In the regression, we use entropy and market excess growth rate (rather than the quantities on the right hand side of \eqref{eqn:diweighted.portfolio}) since they were discussed in depth in previous sections and already give good fit.} This theoretical decomposition does not hold exactly in practice since our data comes from trading subject to transaction costs in a market with dividends, defaults, and a changing set of constituent securities. Nevertheless, 
the model we posit in \eqref{eqn:reg.model} does an excellent job of explaining relative performance. The estimated values of $\beta_1$ and $\beta_2$ are positive, meaning the rebalanced portfolio benefits from a positive change in market diversity and large intrinsic volatility. From the values of these coefficients (and the typical magnitudes of $\Delta \log {\bf H}$ and $\Delta \Gamma_{\mu}$), we also observe that over a short horizon (say a year), a change in market diversity has a larger impact on relative performance than intrinsic volatility. The estimated value of $\beta_0$ is negative; this can be attributed to the increase in transaction costs (relative to the benchmark). Moreover, modulo changes to the estimated coefficients, these conclusions are robust to variations in parameters of the portfolio $w^{p,f}$ and benchmark $w^{1,f}$ which underscores the conceptual importance of these explanatory variables.

\subsubsection{Sensitivity to portfolio specifications}

\begin{figure}[!h]
\includegraphics[scale = 0.64]{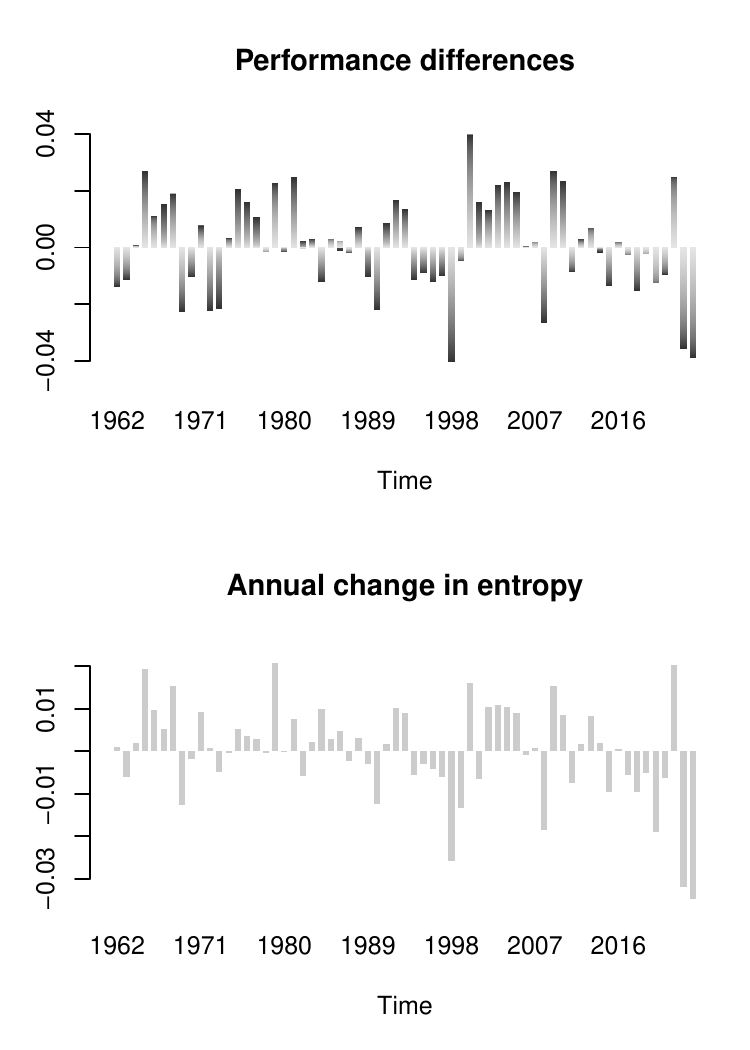}
\includegraphics[scale = 0.64]{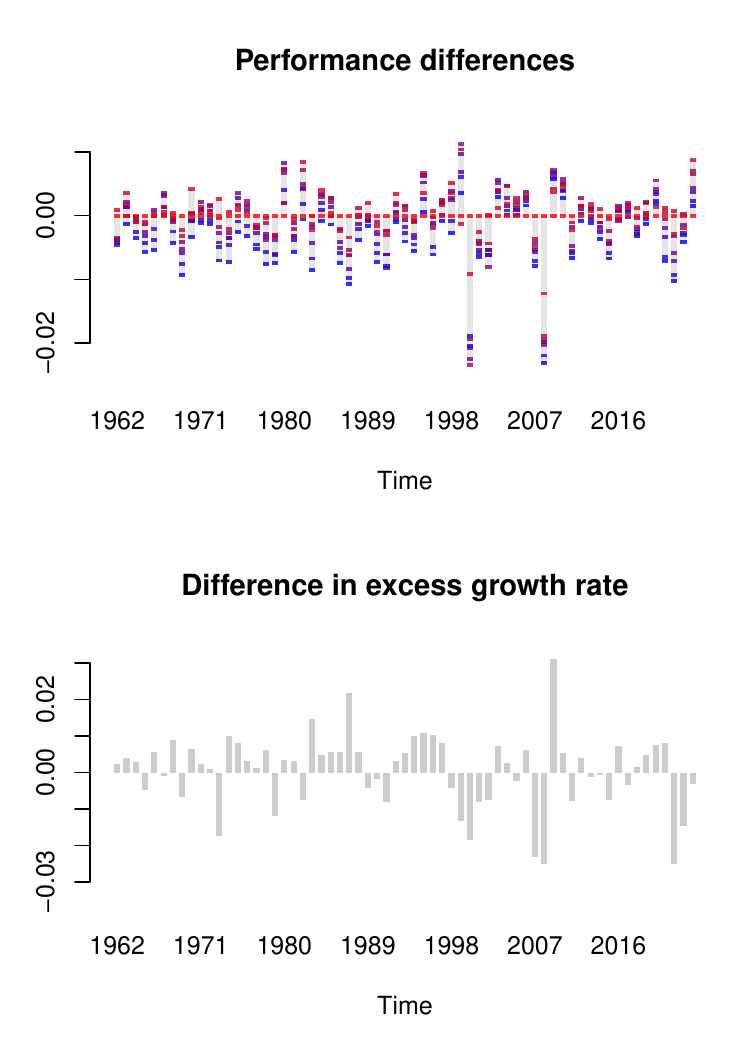}
\vspace{-0.3cm}
\caption{Top Left: Bar plot of $\Delta \log(Z_{p,20}/Z_{1,20})$ for each year on a gradient from black to white as $p$ ranges from $0$ to $1$. Bottom Left: Annual change in entropy, $\Delta\log(\mathbf{H})$. Top Right: Bar plot of $\Delta \log(Z_{0,f}/Z_{0,\infty})$ for each year. The gray bar represents the full range of performance associated with frequencies $f\in\{1,2,5,10,25,50,125,\infty\}$. Solid lines ranging from blue to red mark the performance corresponding to a particular $f$ as the values increase. Bottom Right: Annual change in the cumulative daily excess growth rate $\Delta \Gamma_\mu$ (i.e. $\Delta t= 1$ in Figure \ref{fig:EGR} (left)) less the annual excess growth rate $\gamma_\mu$ \eqref{eqn:excess.growth.rate} for the same period.}
\label{fig:perf.diff}
\end{figure}

Building off of the analysis in Figure \ref{fig:combined.reg}, we can investigate the range of performance that was realised historically by the portfolios $w^{p,f}$. Figure \ref{fig:perf.diff} illustrates the annual performance accessible by varying $p$ and $f$ side-by-side with the explanatory variables from \eqref{eqn:reg.model}. 

The top left panel of Figure \ref{fig:perf.diff} illustrates the annual performance of the (approximately) monthly rebalanced diversity $p$ portfolio, $w^{p,20}$, relative to the index tracking benchmark updated at the same frequency, $w^{1,20}$. Each year has a bar which spans the range of performance for values $p\in[0,1]$. The relative performance of a fixed $p$ is given by the color gradient in the bar which goes from black to white as $p$ increases from $0$ to $1$. By definition, when $p=1$ the relative performance is $0$. In most years, the over/under-performance of the portfolio $w^{p,20}$ is monotone in $p$ as underscored by the monotone change in the bars from black to white. The bottom left panel illustrates the change in entropy over the same calendar years and allows us to compare the performance statistics. It is visually apparent that the directional change in entropy tracks with the orientation of the bars in the top left panel. When entropy increases, smaller values of $p$ tend to outperform and vice-versa.

The top right panel of Figure \ref{fig:perf.diff} plots the annual performance of the equal weight portfolio $w^{0,f}$ relative to the buy-and-hold (initially equal-weighted) portfolio $w^{0,\infty}$ for various {\it fixed} rebalancing frequencies, $f$. Since there are approximately 250 trading days every year, we choose the set of divisors $\{1,2,5,10,25,50,125,\infty\}$\footnote{If a year has 250 trading days then $f=\infty$ will give the same result as any $f>250$.}. The range of the gray bars represents the relative performance spanned by the portfolios in this set. Once again, by construction, when $f=\infty$ the relative performance is $0$. Within each of the bars we mark the value corresponding to a given $f$ by a colored line that changes on a gradient from blue to red as $f$ increases. Unlike the top left panel, the relationship here is not monotone in general. There are many years when intermediate values of $f$ outperform.  We see that, for most years, the choice of frequency is less material than the choice of $p$ in determining the degree of over/under-performance. That said, we observe that frequent rebalancing (say, every 1--2 days) underperforms in most years under $0.25\%$ transaction cost. Also, around the financial crises of 2000 and 2008 the right rebalancing decision appears to meaningfully impact portfolio returns. The bottom right panel displays the difference in the cumulative daily excess growth rate and the annual excess growth rate for the calendar years in our backtest. This statistic captures the difference in the intrinsic volatility at these two different rebalancing scales. However, we see that this only has a weak positive relationship with the difference in performance due to changes in $f$. This is not unexpected, since the contribution of the excess growth rate to portfolio returns is a second order effect on short time horizons (see, e.g.~Figure \ref{fig:combined.reg}) and the strength of the relationship can be confounded by many other market factors including transaction costs, defaults, and a variable set of constituent securities. %

\begin{figure}[!h]
\includegraphics[scale=0.42]{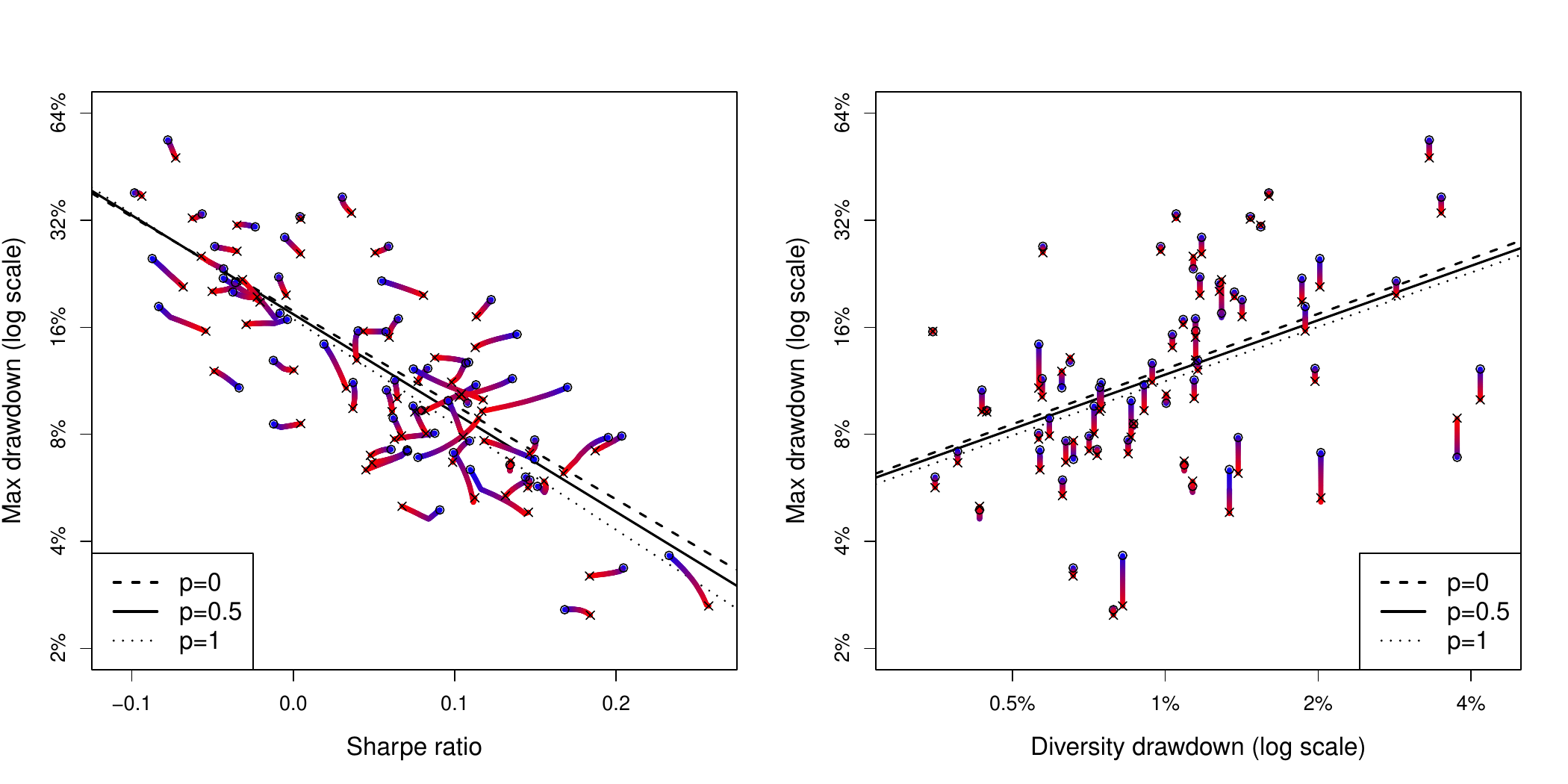}
\vspace{-0.3cm}
\caption{Scatter plots for the risk statistics of the portfolio $w^{p,20}$ (roughly monthly rebalanced). Each point corresponds to a year from 1962 to 2023 on color gradient from blue to red as $p$ ranges from $0$ to $1$. The values corresponding to $p=0$ and $p=1$ are emphasized with the symbols $\circ$ and $\times$, respectively. Left: Annual portfolio drawdown against its Sharpe ratio. Right: Annual portfolio drawdown against the drawdown of market diversity in the same period. In both plots there are lines from a least squares fit of the data for $p\in\{0,0.5,1\}$.}
\label{fig:risk.analysis}
\end{figure}

\subsubsection{Sharpe ratio and drawdown}

Relatively little work has examined portfolio risks and risk-adjusted returns within the framework of stochastic portfolio theory. In their empirical analysis, \cite{RX20} reported the historical Sharpe ratios of several functionally generated portfolios in SPT and illustrated an approximately linear decay in increasing transaction costs. Related theoretical studies include drawdown-constrained growth optimization (e.g.~ \cite{cvitanic1994portfolio} and \cite[Section 4.1]{KK21book}). More recently, the first and last author proposed in \cite{CW22} an optimization framework that allows for flexible exposure to market diversity. In an effort to address this gap, we briefly report here some of the risk characteristics of the diversity-weighted portfolios $w^{p,f}$, which have been central to our empirical study.

Figure \ref{fig:risk.analysis} displays scatter plots of risk statistics for the (approximately) monthly rebalanced diversity portfolios $w^{p,20}$. The left panel shows each calendar year's Sharpe ratio (based on daily returns) versus the maximum drawdown. The right panel compares portfolio drawdowns to drawdowns in market diversity (entropy). In both plots, color indicates the parameter $p\in[0,1]$, and regression lines highlight trends for $p\in\{0,0.5,1\}$.

The results indicate a general inverse relationship between Sharpe ratio and drawdown—years with higher Sharpe ratios tend to experience smaller drawdowns. Portfolio and diversity drawdowns are positively correlated, though the portfolio drawdowns are larger in absolute terms (cf. relative returns in Figure \ref{fig:combined.reg}). Notably, both Sharpe ratios and drawdowns exhibit non-linear dependence on $p$, with some years showing little sensitivity to 
$p$, and others showing substantial variation. On average, increasing $p$ slightly reduces the magnitude of drawdowns and weakens sensitivity to diversity drawdowns. This aligns with the fact that smaller $p$ values overweight smaller-cap stocks, leading to higher turnover and greater portfolio aggressiveness. A deeper understanding of the risk properties of functionally generated portfolios remains an important direction for future research.

\subsection{Portfolio optimization in SPT} \label{sec:portfolio.optimization.literature}
Although we do not address the optimization of large equity portfolios in this paper, we do attempt to briefly survey existing work in this direction. The present literature within SPT can be placed roughly into three groups according to their focus on theory or applications. The largest body of work in this area is theoretical, with the earliest works in SPT focusing on the optimization of outperformance/arbitrage relative to the market \cite{fernholz2010optimal,fernholz2010probabilistic, fernholz2011optimal}. Subsequently, several authors have studied in depth the robust optimization of long run portfolio growth \cite{bayraktar2013robust,itkin2022ergodic,itkin2022robust,IL21,kardaras2012robust,kardaras2021ergodic}. Optimization of functionally generated portfolios in discrete time was considered in \cite{W15} and the works \cite{cuchiero2019cover,wong2015universal}  extended Cover's universal portfolio \cite{cover1991universal} to the setting of SPT, and studied its asymptotic optimality. On the data-driven side, there have been investigations on methods to improve returns by dynamically switching to the market portfolio \cite{TM21} and machine learning approaches to portfolio optimization \cite{SV16}. Living somewhere in the middle, the papers \cite{CW22,CM23} present theory related to classes of portfolio optimization problems, and validate their performance using historical financial data. Specifically, the paper \cite{CW22} studies a non-parametric optimization problem over a special family of rank-based functionally generated portfolios, while \cite{CM23} studies the optimization of linear path-functional portfolios which they prove have a universal approximation property.

\section{Conclusion} \label{sec:conclusion}
In this paper, we conducted a systematic empirical study of several macroscopic properties of the US equity market using the CRSP Database. In the process, we illustrated and highlighted several stylized facts, both old and new. Of particular interest is the apparent importance of entrances and exits in producing a stable market ecosystem. 
Throughout our analysis we have made a concerted effort to suggest novel research directions motivated by stochastic portfolio theory and our findings, and to point to related work that we are aware of where applicable. All of our analysis can be replicated by the code from our repository and we encourage interested readers to interact with the study. In particular, the backtesting engine provided therein may be of independent interest to researchers and practitioners. Apart from the problems discussed throughout the paper, we highlight the following directions for further investigation:
\begin{itemize}
\item[(i)] (Other markets) Most empirical studies in stochastic portfolio theory considered the US equity market largely thanks to the convenient CRSP Database. 
\begin{itemize}
    \item It is interesting to see to what extent the stylized facts we discussed extend to other equity markets in the past and the present. Also, one may consider several markets (e.g.~all Asian markets) as a whole. While we expect many properties (such as the stability of the capital distribution curve and the intensity of rank switching) are common across markets, there are also important differences. To give an example, consider the {\it MSCI Korea Index} \cite{MSCI} which covers about $85\%$ of the Korean equity universe. As of July 2024, the biggest company, Samsung Electronics, covers alone over $30\%$ of the index. In this sense, the Korean market is much more concentrated than the US market.\footnote{For comparision, we note that as of July 2024, the largest three stocks in the S\&P 500 are Microsoft (7.0\%), Apple (6.9\%) and Nvidia (6.2\%).}
    \item The perspective afforded by stochastic portfolio theory can also be applied to non-equity markets. The recent paper \cite{FF22} has already demonstrated a successful application to commodity futures markets. A study of alternative markets from a macroscopic lens would be interesting due to their varied properties. For instance, these commodity markets do not have ``entrances" and ``exits" in the traditional sense.
\end{itemize}
\item[(ii)] (Other quantities and intraday data) In this paper we focused on macroscopic properties defined in terms of market capitalizations and returns. Other quantities, such as trading volumes, signatures (in the sense of rough path theory) \cite{CGS23}, sectors, prices of options\footnote{See \cite{fernholz2023portfolios} for some relationships between contingent claims and functionally generated portfolios in SPT.} and exchange traded funds, may offer additional insights. Also, the CRSP Database offers daily but not intraday data. SPT-inspired analysis of high frequency data is promising but largely open; the only work we are aware of is \cite{F07} which studied intraday properties of the excess growth rate in relation to the profits made by market makers. A possible direction is to study intraday return distributions (see e.g.~\cite{ZKP22}) and relate them with market capitalizations and other quantities.
\item[(iii)] (Adaptive portfolio selection under realistic conditions) Our empirical study and the backtest conducted in Section \ref{sec:backtest} naturally suggest the study of systematic approaches for portfolio selection with a large but evolving number of assets and under transaction costs. Short positions, derivatives, and adaptive rebalancing frequencies may also be considered. Needless to say, these and other issues are involved in all actual portfolios but are not easy to incorporate in theoretical models in mathematical finance. We hope our work serves as a bridge between the theoretical and practical sides.
\end{itemize}

\section*{Acknowledgement}
S.~Campbell and T.-K.~L.~Wong thank Martin Larsson, David Itkin and Robert Jones for helpful discussions. We also thank the anonymous reviewers for their careful reading and insightful comments. 
The research of T.-K.~L.~Wong is partially supported by an NSERC Discovery Grant (RGPIN-2019-04419) and a Seed Funding for Methodologists Grant from the Data Sciences Institute (DSI) at the University of Toronto. The authors report there are no competing interests to declare.

\appendix
\section*{Appendix}
\section{Crash course in stochastic portfolio theory} \label{sec:appendix.spt}
We provide an overview of some basic concepts of {\it stochastic portfolio theory} (SPT). Our aim is to describe an idealized market model and use it to motivate some of the topics discussed in the main text. An especially important result is the identity \eqref{eqn:diweighted.portfolio} which is relevant to our discussions of market diversity in Section \ref{sec:diversity} and portfolio backtesting in Section \ref{sec:backtest}. It is a special case of the so-called {\it master formula} in SPT. The master formula illustrates that the relative performance of certain systematically rebalanced portfolios can be decomposed in terms of the change in market diversity and a quantity which measures cumulative relative volatility. Systematic introductions to SPT, including the master formula,
can be found in standard references of the subject such as \cite{F02, FK09} (also see \cite{KK21book}). We also refer the reader to \cite{PW16, W19} for a discrete time formulation which does not require stochastic calculus. Another model-free approach using {\it rough paths} in continuous time is developed in \cite{ACLP21}. 

Consider an equity market model consisting of a {\it fixed} collection of $n$ non-dividend paying stocks.\footnote{Dividends are included in Fernholz's original market model \cite{F02} but are frequently omitted for convenience. See \cite{BKT23, KK21} for recent attempts to relax the assumption of a fixed investment universe.} The market capitalizations $X_1(t), \ldots, X_n(t)$ of the stocks are modelled as a vector of positive It\^{o} processes defined on some filtered probability space. The vector $\mu(t) = (\mu_1(t), \ldots, \mu_n(t))$ of market weights, where $\mu_i(t) = X_i(t)/(X_1(t) + \cdots + X_n(t))$, takes values in the open unit simplex $\Delta_n = \{ \mu = (\mu_1, \ldots, \mu_n) \in (0, 1)^n: \mu_1 + \cdots + \mu_n = 1\}$. The {\it diversity} of the market refers to the concentration of the capital distribution, and may be quantified in terms of a symmetric concave function $\Phi: \Delta_n \rightarrow (0, \infty)$ (see \cite[Section 3.4]{F02}). Popular choices include the {\it Shannon entropy} ${\bf H}(\mu)$ (see \eqref{eqn:Shannon.entropy}) as well as the parameterized diversity ${\bf D}_{p}(\mu)$ (see \eqref{eqn:Fernholz.diversity.def}). It was observed empirically that market diversity tends to ``mean-reverting'' in the long run (see Section \ref{sec:diversity}). At the very least, the market weight vector $\mu(t)$ tends to avoid certain regions of the simplex especially its vertices. Motivated by this and other empirical observations, various {\it path properties} may be imposed on the market model. For example, the market is said to be {\it coherent} if a.s.~$\lim_{t \rightarrow \infty} \frac{1}{t} \log \mu_i(t) = 0$ for all $i$, and is {\it diverse} if there exists $\delta > 0$ such that a.s.~$\max_{1 \leq i \leq n} \mu_i(t) \leq 1 - \delta$ for all $i$ and $t \geq 0$. Note that some conditions, when imposed to hold with probability $1$, exclude the existence of an equivalent martingale measure (over a finite horizon); see \cite[Section 6]{FK09} for a discussion.

An {\it all-long self-financing portfolio} is represented by a progressively measurable process $\pi(t) = (\pi_1(t), \ldots, \pi_n(t))$ with values in the closure of $\Delta_n$. We interpret $\pi_i(t)$ as the percentage of current capital invested in stock $i$. Henceforth all portfolios are assumed to be all-long and self-financed. In SPT, we focus on the {\it relative value} $V_{\pi}(t)$ of the portfolio with respect to the market portfolio. This amounts to taking the value $X_1(t) + \cdots + X_n(t)$ of the market portfolio as the numeraire. Assuming frictionless continuous trading, the relative value satisfies $\frac{\dd V_{\pi}(t)}{V_{\pi}(t)} = \sum_{i = 1}^n \pi_i(t) \frac{\dd \mu_i(t)}{\mu_i(t)}$. By It\^{o}'s formula, we have
\begin{equation} \label{eqn:portfolio.basic.decomp}
\dd \log V_{\pi}(t) = \sum_{i = 1}^n \pi_i(t) \dd \log \mu_i(t) + \dd \Gamma_{\pi}(t),
\end{equation}
where $\Gamma_{\pi}(t)$ is a non-decreasing finite variation process called the {\it cumulative excess growth rate} of the portfolio. It is given by
\begin{equation} \label{eqn:dGamma}
\dd \Gamma_{\pi}(t) = \frac{1}{2} \left( \sum_{i = 1}^n \pi_i(t) \dd \langle \log \mu_i \rangle (t)  - \sum_{i, j = 1}^n \pi_i(t) \pi_j(t) \dd \langle \log \mu_i, \log \mu_j \rangle(t) \right),
\end{equation}
where $\langle \cdot \rangle$ denotes quadratic (co)variation. Its increments can be approximated by the (discrete) {\it excess growth rate} (see Section \ref{sec:excess.growth.rate}): for $\Delta t > 0$ small we have
\begin{equation} \label{eqn:excess.growth.rate.approximation}
\Gamma_{\pi}(t + \Delta t) - \Gamma_{\pi}(t) \approx \log \left( \sum_{i = 1}^n \pi_i(t) \frac{\mu_i(t + \Delta t)}{\mu_i(t)} \right) - \sum_{i = 1}^n \pi_i(t) \log \frac{\mu_i(t + \Delta t)}{\mu_i(t)}.
\end{equation}
The cumulative excess growth rate $\Gamma_{\mu}(t)$ of the {\it market portfolio} $\pi \equiv \mu$ is especially important and may be regarded as a measure of the {\it intrinsic} (or {\it relative}) {\it volatility} of the market. This concept decouples the absolute volatility of the market with the relative volatility among the constituent stocks. For the market portfolio, \eqref{eqn:dGamma} simplifies and yields
\begin{equation} \label{eqn:market.excess.growth.rate}
\dd \Gamma_{\mu}(t) = \frac{1}{2} \sum_{i = 1}^n \mu_i(t) \dd \langle \log \mu_i \rangle (t) = \frac{1}{2} \sum_{i = 1}^n \frac{\dd \langle \mu_i \rangle(t)}{\mu_i(t)}.
\end{equation}
In fact, $\Gamma_{\mu}(t)$ is ($\frac{1}{2}$ times)  the Riemannian quadratic variation of the simplex-valued continuous semimartingale $\mu(\cdot)$ if we equip $\Delta_n$ with the {\it Fisher-Rao Riemannian metric} (see \cite{A16, E89} for the precise definitions). In \cite{FKR18}, $\Gamma_{\mu}(t)$ is interpreted as a measure of intrinsic market volatility.

A portfolio $\pi$ is said to be a {\it relative arbitrage} with respect to the market portfolio over a finite horizon $[0, T]$, if
\begin{equation} \label{eqn:relative.arbitrage}
\mathbb{P}\left( V_{\pi}(T) \geq V_{\pi}(0)\right) = 1 \quad \text{and} \quad \mathbb{P}\left( V_{\pi}(T) > V_{\pi}(0)\right) > 0.
\end{equation}
A key insight of SPT is that relative arbitrages exist under fairly realistic conditions which do not fully specify the market model. In this , SPT is partially {\it model-free}.

To illustrate this idea, we first consider a {\it constant-weighted portfolio} $\pi(t) \equiv \pi \in \Delta_n$ that continuously rebalances to the same weight. From \eqref{eqn:portfolio.basic.decomp}, its relative growth rate satisfies
\begin{equation} \label{eqn:constant.weighted.decomp}
\frac{1}{t} \log \frac{V_{\pi}(t)}{V_{\pi}(0)} = \sum_{i = 1}^n \frac{1}{t} \pi_i \log \frac{\mu_i(t)}{\mu_i(0)} + \frac{1}{t} \Gamma_{\pi}(t).
\end{equation}
If the market is coherent, the first term on the right hand side of \eqref{eqn:constant.weighted.decomp} tends to $0$ as $t \rightarrow \infty$. Suppose further that the market is diverse and is {\it nondegenerate} in the sense that the covariance process of $\log X(t)$ is uniformly elliptic \cite[Definition 1.1.2]{F02}. Then, it can be shown that $\dd \Gamma_{\pi}(t) \geq \epsilon \dd t$ for some constant $\epsilon > 0$. Letting $t \rightarrow \infty$ in \eqref{eqn:constant.weighted.decomp}, we have $\liminf_{t \rightarrow \infty} \frac{1}{t} \log \frac{V_{\pi}(t)}{V_{\pi}(0)} \geq \epsilon$ a.s., thus giving an ``asymptotic relative arbitrage''.

To give an explicit example of relative arbitrage in the sense of \eqref{eqn:relative.arbitrage}, consider the {\it diversity-weighted portfolio} which is closely related to the function ${\bf D}_p$ in \eqref{eqn:Fernholz.diversity.def} and is given by
\begin{equation} \label{eqn:diversity.weighted.portfolio}
\pi_i(t) = \frac{\mu_i^{p}(t)}{\sum_{j = 1}^n \mu_j^{p}(t)}, \quad i = 1, \ldots, n,
\end{equation}
where $p \in (0, 1)$ is a tuning parameter. Note that letting $p \rightarrow 0$ recovers the equal-weighted portfolio $\pi_i(t) \equiv \frac{1}{n}$ and letting $p \rightarrow 1$ recovers the market portfolio $\pi(t) \equiv \mu(t)$. Using It\^{o}'s formula, one can show that the relative value of the portfolio satisfies the {\it pathwise decomposition} or the {\it master formula}
\begin{equation} \label{eqn:diweighted.portfolio}
\log \frac{V_{\pi}(t)}{V_{\pi}(0)} = \log \frac{{\bf D}_{p}(\mu(t))}{{\bf D}_{p}(\mu(0))} + (1 - p) \Gamma_{\pi}(t), \quad t \geq 0.
\end{equation}
From \eqref{eqn:diversity.weighted.portfolio}, the relative performance of the portfolio is characterized by (i) the change in market diversity and (ii) market volatility measured by the excess growth rate of the portfolio. The excess growth rate  contributes positively to relative performance, but over the short run the dynamics of $\log V_{\pi}(t)$ is dominated by that of $\log {\bf D}_p(\mu(t))$. As the parameter $p$ varies between $0$ and $1$, we have a trade-off between exposure to intrinsic volatility versus diversity.\footnote{See \cite{vervuurt2015diversity} for an analysis of the diversity-weighted portfolio when $p$ is {\it negative}.} In practice, a master formula such as \eqref{eqn:diweighted.portfolio} only holds approximately due to complications including transaction costs, dividends and corporate actions. Nevertheless, it identifies market diversity and relative volatility as two major drivers of the relative value of the portfolio.

To complete the construction of a relative arbitrage, assume that the market is diverse, so that $\log {\bf D}_{p}(\mu(t)) \geq -M$ for some $M > 0$. If the market is also nondegenerate, then $\dd (1 - p) \Gamma_{\pi}(t) \geq \epsilon \dd t$ for some $\epsilon > 0$. This gives the a.s.~lower bound
\[
\log \frac{V_{\pi}(t)}{V_{\pi}(0)} \geq -M - \log {\bf D}_{p}(\mu(0)) + \epsilon t,\quad t \geq 0,
\]
and we have a relative arbitrage over $[0, T]$ whenever $T > T^* := \frac{1}{\epsilon}(M + \log {\bf D}_{p}(\mu(0)))$ (unfortunately, the bound $T^*$ obtained this way is too large for practical purposes). Both the constant-weighted portfolio and the diversity-weighted portfolio are special cases of {\it functionally generated portfolios} for which a pathwise decomposition analogous to \eqref{eqn:diversity.weighted.portfolio} holds.\footnote{Analogous pathwise decompositions can be established without a stochastic model. See \cite{PW16} for the discrete time case and \cite{ACLP21} for the continuous time case using rough path theory. For generalizations of functional portfolio generalization see \cite{karatzas2017trading, kim2023market, W19} and the references therein.} Relative arbitrages can also be constructed over short time horizons or under weaker conditions. For further details and more recent developments, we refer the reader to \cite{ACLP21, CW22, FKR18, KK21, PW16} and the references therein. 

\section{Local time} \label{sec:details}
We recall the concept of {\it semimartingale local time} in stochastic calculus which motivates the development in Section \ref{sec:local.time.empirical} and is closely related to the concept of leakage in stochastic portfolio theory. We follow the notations of \cite[Section 4.1]{F02}. Further details can be found in \cite[Chapter VI]{revuz2013continuous}, \cite{BG08} and \cite{X20}. Let $Y(t)$ be a real-valued continuous semimartingale defined on a filtered probability space. The {\it semimartingale local time} of $Y$ at $0$ is the non-decreasing process $\Lambda$ defined by the identity
\begin{equation} \label{eqn:local.time}
\frac{1}{2} |Y(t)|  =  \frac{1}{2}|Y(0)| + \frac{1}{2} \int_0^t \sign(Y(s)) \dd Y(s) + \Lambda(t),
\end{equation}
where $\sign(y) = 1$ if $y > 0$ and $\sign(y) = -1$ if $y \leq 0$. We may regard \eqref{eqn:local.time} as an extension of It\^{o}'s formula applied to the convex function $y \mapsto \frac{1}{2}|y|$ whose distributional second derivative is a point mass at $0$. It can be shown that $\dd \Lambda(t) = 0$ on $\{t : Y(t) \neq 0\}$. Intuitively, $\Lambda(t)$ characterizes the ``amount of time'' $Y$ spends at the point $0$ (this is justified by the {\it occupation formula} which involves the local time at each $y \in \mathbb{R}$).

Next consider an $n$-dimensional continuous semimartingale $(Y_1,\dots,Y_n)$. We say that $Y_1,\dots,Y_n$ are \textit{pathwise mutually non-degenerate} if:
\begin{enumerate}
    \item[(i)] for all $i\not=j$, $\{t:Y_i(t)=Y_j(t)\}$ has Lebesgue measure zero a.s.;
    \item[(ii)] (triple collision) for all $i<j<k$, $\{t: Y_i(t)=Y_j(t)=Y_k(t)\}=\emptyset$ a.s.
\end{enumerate}
Write $Y_i(t) = Y_i(0) + M_i(t) + A_i(t)$ where $M_i$ is the local martingale part and $A_i$ has finite variation. We say that $Y_i$ is \textit{absolutely continuous} if the random signed measures $A_i$ and $\langle M \rangle_i$ are a.s.~absolutely continuous with respect to the Lebesgue measure. Let $Y_{(1)}(t) \geq \cdots \geq Y_{(n)}(t)$ be the non-increasing order statistics of $Y_1(t), \ldots, Y_n(t)$. We are interested in the {\it gaps} $Y_{(1)} - Y_{(2)}, \ldots, Y_{(n-1)} - Y_{(n)}$ which are non-negative by construction. Since $Y$ is continuous, if a rank switch between ranks $k$ and $k + 1$ occurs at $t$ then $Y_{(k)}(t) - Y_{(k+1)}(t) = 0$. Thus we may use the local time of $Y_{(k)} - Y_{(k+1)}$ at $0$ to measure to intensity of rank switching.

\begin{proposition}\label{prop:rank.dynamics}\cite[Proposition 4.1.11]{F02} Let $Y_1,\dots,Y_n$ be pathwise mutually non-degenerate and absolutely continuous. Then $(Y_{(1)}, \ldots, Y_{(n)})$ is an $n$-dimensional continuous semimartingale. For $k = 1, \ldots, n - 1$, let $\Lambda_{k}(t)$ be the local time of the non-negative semimartingale $Y_{(k)} - Y_{(k+1)}$ at $0$. Also let $\Lambda_{0}(t) = \Lambda_{n+1}(t) = 0$. Then
\begin{equation} \label{eqn:ranked.process.SDE}
\dd Y_{(k)}(t)=\sum_{i=1}^n\mathds{1}_{\{Y_{i}(t)=Y_{(k)}(t)\}}\dd Y_i(t)+ \frac{1}{2}\left(\dd\Lambda_{k}(t)-\dd\Lambda_{k-1}(t)\right).
\end{equation}
\end{proposition}
This result enables one to recover the local times of the gaps $Y_{(k)} - Y_{(k+1)}$ from the ranked processes $Y_{(k)}(t)$ and stochastic integrals of the form $\sum_i \mathds{1}_{\{Y_{i}(t)=Y_{(k)}(t)\}}\dd Y_i(t)$. In the context of Section \ref{sec:local.time.empirical}, $Y_i(t) = \log X_i(t)$ is the log market capitalization of stock $i$. We remark here that the case when (i) and (ii) do not hold is studied by \cite{BG08}. Dynamics of the type in \eqref{eqn:ranked.process.SDE} can be derived, but their expression is considerably more complicated.

\bibliographystyle{abbrv}
\bibliography{geometry.ref}

\end{document}